\documentclass[12pt,a4paper]{article}

\usepackage{graphics}
\usepackage{latexsym}
\usepackage{amsmath}
\usepackage{amssymb}
\usepackage{slashed}
\usepackage[usenames]{color}

\title{Are James Webb Space Telescope observations consistent with
warm dark matter?}
\author{Bruce Hoeneisen}
\date{\small{
Universidad San Francisco de Quito, Quito, Ecuador \\
Email: bhoeneisen@usfq.edu.ec \\
16 January 2024}
}

\begin{document}
\maketitle

\begin{abstract}
\noindent
We compare observed with predicted distributions of galaxy stellar masses
$M_*$ and galaxy rest-frame ultra-violet luminosities per unit bandwidth $L_{UV}$,
in the redshift	range $z = 2$ to 13.
The comparison is presented as a function of the comoving warm dark matter
free-streaming cut-off wavenumber $k_{fs}$.
For this comparison the theory is a minimal extension of the
Press-Schechter	formalism with only two	parameters:
the star formation efficiency, and a proportionality factor
between	the star formation rate per galaxy and $L_{UV}$.
These two parameters are fixed to their	values obtained	prior
to the James Webb Space	Telescope (JWST) data. The purpose
of this	comparison
is to identify if, and where, detailed astrophysical evolution
is needed to account for the new JWST observations.
\end{abstract}

\noindent
\textbf{Keywords} \\
James Webb Space Telescope, JWST, Warm Dark Matter.

\section{Introduction}
\label{introduction}

Early James Webb Space Telescope (JWST) observations 
were surprising: they reveal
galaxies at high redshifts ($z \gtrsim 10$) with greater number densities
than predicted by the ``$\Lambda$CDM cosmology".
A list of surprises form the first observations of JWST, and
of several proposed modifications of the theory to
account for this data, is summarized in \cite{Gupta}.
The ``$\Lambda$CDM cosmology" that is actually compared with
the new JWST data is an extension of the 6-parameter $\Lambda$CDM 
cosmology with a dozen, or so,
astrophysical parameters \cite{Behroozi}.

The purpose of the present study is two-fold:
a) To extend the comparison between observations and predictions
to include warm	dark matter, and
b) to make the comparison with a ``first-order"	prediction, that
has as few, physically motivated, adjustable parameters as possible, 
with these parameters determined numerically from observations prior to
JWST data, leaving zero	new degrees of freedom for the comparison.
The idea is to clearly identify	regions	of parameter space that
require a more detailed cosmological and/or astrophysical description 
to understand the evolution of the observations.

We focus our attention on distributions of the galaxy stellar masses $M_*$ and of the
galaxy rest-frame ultra-violet (UV) luminosities per unit bandwidth $L_{UV}$, that can be
predicted with the Press-Schechter formalism
given three parameters: the warm dark matter power spectrum cut-off wavenumber $k_{fs}$,
the stellar formation efficiency $f_*$, and a proportionality factor
between the star formation rate per galaxy (SFR) and $L_{UV}$.

The outline of the article is as follows. In Section \ref{comparison}
we compare observed and predicted distributions of $M_*$ and
$L_{UV}$ for redshifts $z$ in the range 2 to 13.
The ``first order" theory turns out to be in agreement with
most of the data. There are, however, three parameter regions
with tensions. In Section \ref{data} we describe the
data. In Section \ref{predictions}  we present the details
of how we include warm dark matter in the Press-Schechter formalism.
With these preparations we are able to discuss the
observed tensions in Section \ref{discussion}. We close with conclusions.

\section{Comparison of observed distributions of $M_*$ and $L_{UV}$ with predictions}
\label{comparison}

\begin{figure}
\begin{center}
%\vspace*{-4.5cm}
\scalebox{0.7}
%{\includegraphics{UV_Press_Schechter_z6_261223_PK.eps}}
%{\includegraphics{UV_Press_Schechter_z6_281223_PK.eps}}
%{\includegraphics{UV_Press_Schechter_z6_311223_PK.eps}}
%{\includegraphics{UV_Press_Schechter_z6_090124_PK.eps}}
{\includegraphics{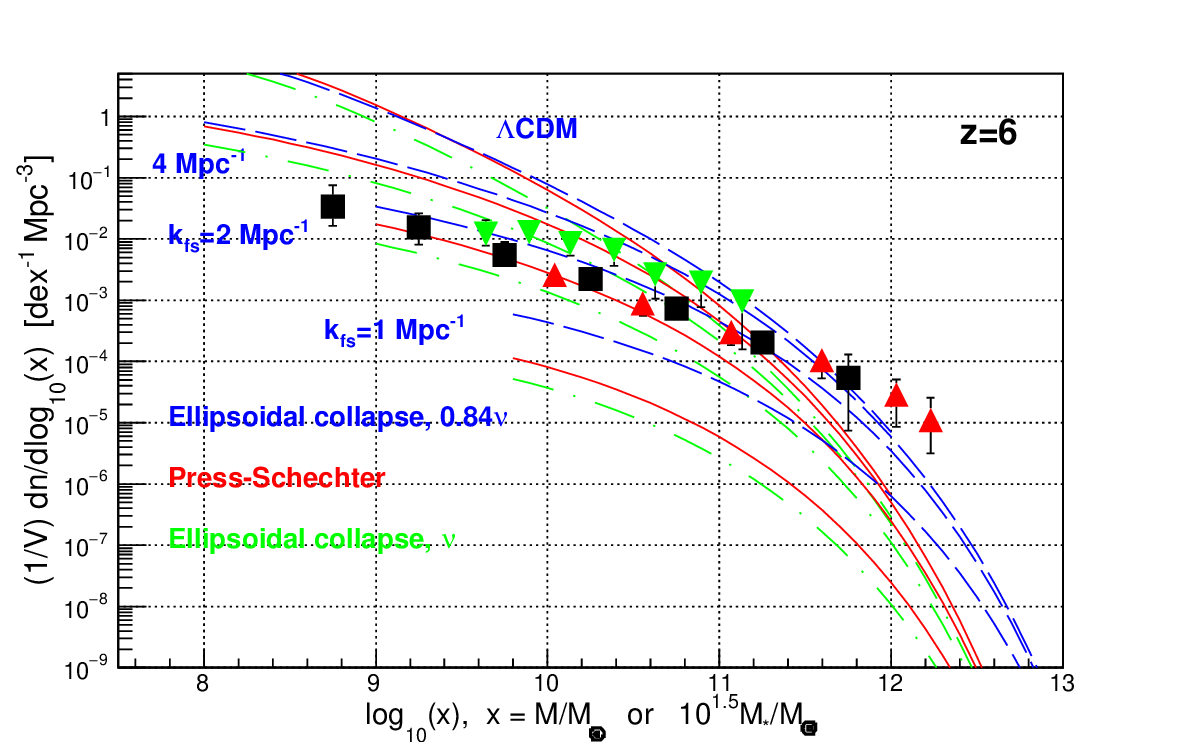}}
\scalebox{0.7}
%{\includegraphics{UV_Press_Schechter_z6_261223_.eps}}
%{\includegraphics{UV_Press_Schechter_z6_281223_.eps}}
%{\includegraphics{UV_Press_Schechter_z6_311223_.eps}}
%{\includegraphics{UV_Press_Schechter_z6_090124_.eps}}
{\includegraphics{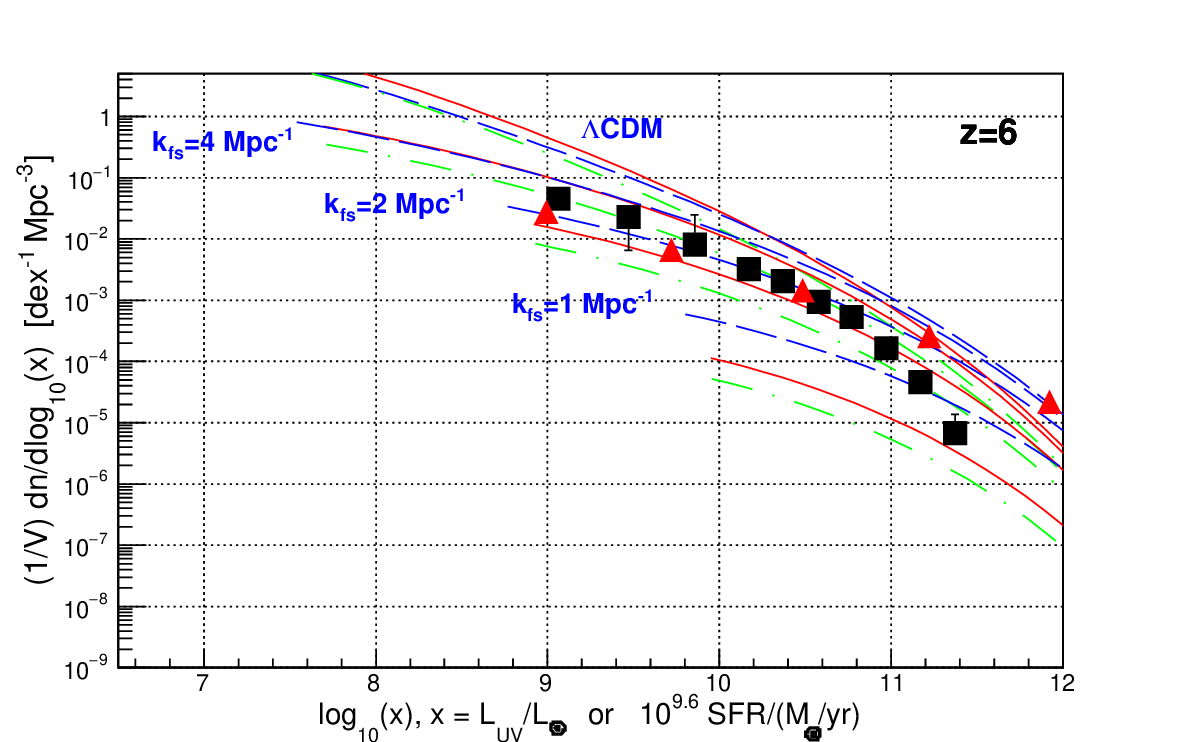}}
%\vspace*{0.7cm}
\caption{Comparison of predicted and observed
distributions of $M/M_\odot = 10^{1.5} M_*/M_\odot$ (top panel) 
and $L_{UV}/L_\odot = 10^{9.6} \textrm{SFR}/(M_\odot/\textrm{yr})$
(bottom panel) for redshift $z = 6$.
Data are from the Hubble Space Telescope ($M_*$ from \cite{Song} and
$L_{UV}$ from \cite{Bouwens}) (black squares), 
from the continuity equation \cite{Lapi} (red triangles), and
from the James Webb Space Telescope (green triangles) \cite{Navarro}.
%https://iopscience.iop.org/article/10.3847/0004-637X/825/1/5/pdf Table 2 \cite{Song}
%https://arxiv.org/pdf/2102.07775.pdf Table 4 \cite{Bouwens}
%https://arxiv.org/pdf/1708.07643.pdf fig. 4 and fig. 1 \cite{Lapi}
}
\label{PS_z6}
%/home/bruce1/JWST
%UV_Press_Schechter.C_bck081223
\end{center}
\end{figure}

In Figure \ref{PS_z6} we compare observed
distributions of galaxy stellar masses $M_*$ (top panel) and
galaxy rest-frame ultra-violet luminosities per unit bandwidth $L_{UV}$ (bottom panel)
with ``first-order" predictions at redshift $z = 6$. The data is obtained from
Hubble Space Telescope (HST) observations (black squares) \cite{Song} \cite{Bouwens},
from the continuity equation \cite{Lapi} (red triangles), and from the
James Webb Space Telescope (JWST) observations (green triangles) \cite{Navarro}.
A description of this data is postponed to Section \ref{data}. 

Experimenters obtain the galaxy absolute magnitude $M_{1600,AB}$ in the AB system \cite{PDG2023}
at the rest frame wavelength $\approx 1600$ \AA \ in the ultra-violet (UV). 
This absolute magnitude is related to the galaxy radiated
power per unit  bandwidth $L_{1600}$
as follows:
\begin{equation}
M_{1600,AB} = +51.6 - 2.5 \log_{10}{L_{UV}},
\label{Mgal}
\end{equation}
where $L_{UV} \equiv L_{1600}/(\textrm{erg s}^{-1} \textrm{ Hz}^{-1})$ (from the definition
of absolute $M_{AB}$ in \cite{PDG2023}).
In comparison, the absolute magnitude of the Sun at wavelength $3900$ \AA \ is \cite{Willmer}
\begin{equation}
M_{3900,\odot,AB} = 5.9 = +51.6 - 2.5 \log_{10}{L_\odot},
\label{Msun}
\end{equation}
where $L_\odot \equiv L_{3900,\odot}/(\textrm{erg s}^{-1} \textrm{ Hz}^{-1})$.
Note that the solar flux power per unit wavelength is a maximum near $3900$ \AA.
We plot distributions of $M_*/M_\odot$ and $L_{UV}/L_\odot$. 
$L_{UV}/L_\odot$ is obtained from $M_{1600,AB}$ as follows:
\begin{equation}
M_{1600,AB} \equiv 5.9 - 2.5 \log_{10} { \left( \frac{L_{UV}}{L_\odot} \right) }.
\end{equation}
This definition of $L_{UV}/L_\odot$, that is somewhat arbitrary, is chosen
because it gives a physical sense of luminosity in solar units, and also because
$M/M_\odot \approx L_{UV}/L_\odot$ at $z = 6$, see Figure \ref{PS_z6}.

The predictions are an extension of
the Press-Schechter formalism to include warm dark matter
(as described in Section \ref{predictions}). The
``warmness" of the dark matter is defined by the 
comoving power spectrum cut-off wavenumber $k_{fs}$ due to
dark matter free-streaming in and out of density 
minimums and maximums. 
We present predictions corresponding to $k_{fs} = 1, 2, 4$ and 200 Mpc$^{-1}$.
The latter large value of $k_{fs}$, corresponding to negligible
free-streaming, is identified with the cold dark matter $\Lambda$CDM cosmology.
We present Press-Schechter predictions \cite{Press-Schechter}, and
two Sheth-Mo-Tormen ellipsoidal collapse extensions \cite{Sheth_Tormen} \cite{Sheth_Mo_Tormen}.
Presenting three predictions for each $k_{fs}$ illustrates the uncertainties of these
predictions. 
The predictions obtain the distributions of
the \textit{linear} perturbation total (dark matter plus baryon) masses $M$
as defined by the Press-Schechter formalism (see Section \ref{predictions}). 
The predicted number of galaxies per unit volume and per decade of $M$,
$(1/V) \cdot dn/d\log_{10}(M/M_\odot)$, with units $[\textrm{dex}^{-1} \textrm{Mpc}^{-3}]$,
is a function of $M$, $z$ and $k_{fs}$.
Therefore it is still
necessary to find relations between $M_*$ and $L_{UV}$ with $M$.
We consider the simplest relations, i.e. $M_*$ proportional to $M$, and
$L_{UV}$ proportional to the star formation rate per galaxy (SFR),
with units $\left[ M_\odot/\textrm{yr}\right]$.
The  proportionality of $L_{UV}$ with SFR is justified
because $L_{UV}$ is dominated by large mass stars with lifetimes $\tau$ short
compared to the age of the universe $t(z)$ at redshift $z$ \cite{Madau}. Then
\begin{eqnarray}
M_* & \equiv & 10^{-a} M \equiv f_* \frac{\Omega_b}{\Omega_c + \Omega_b} M, \label{Mstar} \\
\frac{L_{UV}}{L_\odot} & \equiv & 10^b \frac{\textrm{SFR}}{M_\odot/\textrm{yr}}, \label{LUV1}
\end{eqnarray}
which define the parameters $a$, $f_*$ and $b$.
$\Omega_b$ and $\Omega_c$ are the mean densities of baryons and dark matter in units of
the critical density (throughout we use the notation and parameter values of \cite{PDG2023}).
$f_*$ is defined to be the ``star formation efficiency".
Given $(M, z, k_{fs})$ we predict
\begin{equation}
\textrm{SFR} \equiv 10^{-a} \frac{M}{\tau}  \frac{dn/d\log_{10}(M/M_\odot) - dn'/d\log_{10}(M/M_\odot)}
{dn/d\log_{10}(M/M_\odot)},
\label{SFR}
\end{equation}
where $dn/d\log_{10}(M/M_\odot)$ is calculated at $(M, z, k_{fs})$, and
$dn'/d\log_{10}(M/M_\odot)$ is calculated at $(M, z', k_{fs})$, where
the age of the universe at redshift $z$ is $t(z)$, and
the age of the universe at redshift $z'$ is $t(z') = t(z) - \tau$,
in the limit of small $\tau$, i.e. $\tau \ll t(z)$.

In summary, the predictions for each $k_{fs}$ depend on only the two parameters
$a$ and $b$. These parameters are in principle functions of $M$, $z$ and $k_{fs}$.
However, for the purpose of comparisons with the data, we assume that $a$ and $b$ are
constants. Note that varying $a$ and $b$ shifts the predictions in Figure \ref{PS_z6}
to the right or to the left. For the comparisons we choose values of $a$ and $b$ obtained
\textit{prior} to JWST data \cite{fermion_or_boson} \cite{LUV}:
\begin{equation}
a = 1.5 \qquad \textrm {and} \qquad b = 9.6.
\label{fa}
\end{equation}
These values of $a$ and $b$ define our ``first-order" predictions.
$a = 1.5$ corresponds to a star formation efficiency $f_* = 0.20$
(taken to be independent of $z$ in our ``first-order" predictions!).
For a Salpeter initial mass function (IMF), and $L_{UV}$ measured at a rest
frame wavelength $\approx 1500 \AA$, the following $L_{UV}$ is obtained in \cite{Madau}:
\begin{equation}
L_{UV} = 8 \times 10^{27} \frac{\textrm{SFR}}{M_\odot/\textrm{yr}}.
\label{Madau}
\end{equation}
Then, from (\ref{Msun}), (\ref{LUV1}) and (\ref{Madau}), we obtain $b = 9.6$.
The value of $k_{fs}$ measured \textit{prior} to JWST data is \cite{LUV}
\begin{equation}
k_{fs} = 2.0^{+0.8}_{-0.5} \textrm{ Mpc}^{-1}.
\end{equation}

\begin{figure}
\begin{center}
%\vspace*{-4.5cm}
\scalebox{0.33}
%{\includegraphics{UV_Press_Schechter_z2_091223_PK.eps}}
%{\includegraphics{UV_Press_Schechter_z2_111223_PK.eps}}
%{\includegraphics{UV_Press_Schechter_z2_211223_PK.eps}}
%{\includegraphics{UV_Press_Schechter_z2_281223_PK.eps}}
{\includegraphics{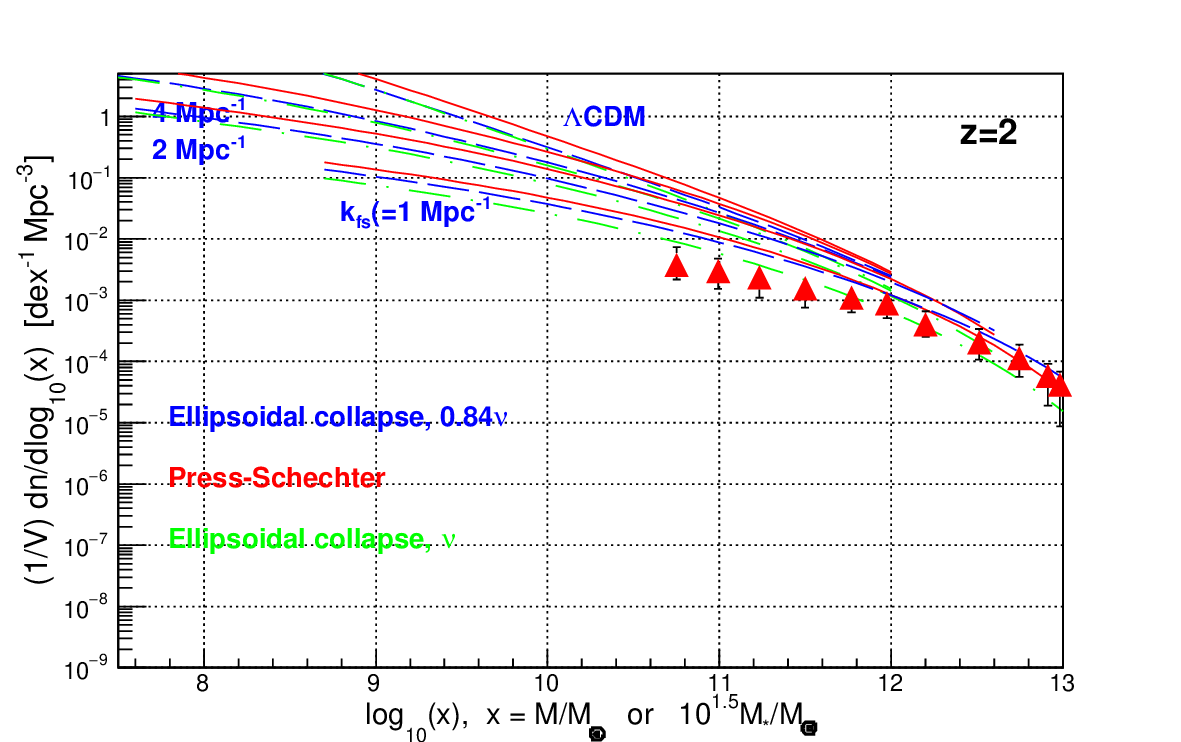}}
\scalebox{0.33}
%{\includegraphics{UV_Press_Schechter_z2_091223_.eps}}
%{\includegraphics{UV_Press_Schechter_z2_111223_.eps}}
%{\includegraphics{UV_Press_Schechter_z2_211223_.eps}}
%{\includegraphics{UV_Press_Schechter_z2_281223_.eps}}
{\includegraphics{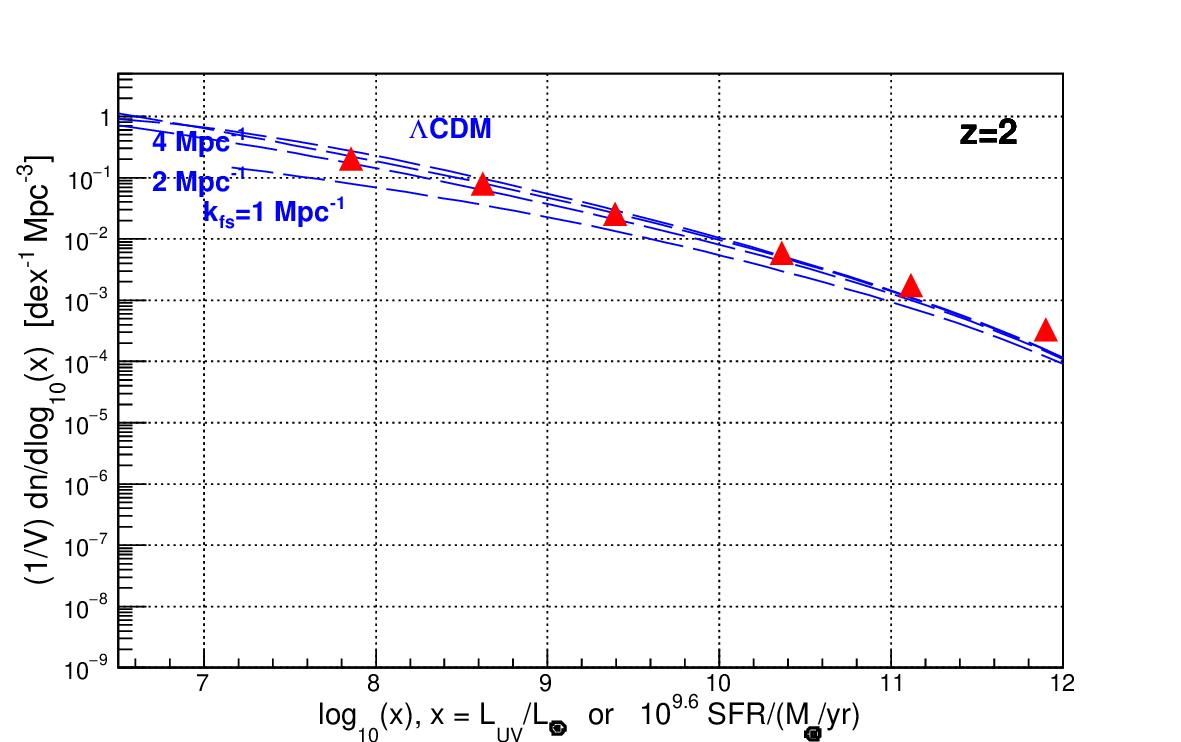}}
\scalebox{0.33}
%{\includegraphics{UV_Press_Schechter_z3_091223_PK.eps}}
%{\includegraphics{UV_Press_Schechter_z3_111223_PK.eps}}
%{\includegraphics{UV_Press_Schechter_z3_211223_PK.eps}}
%{\includegraphics{UV_Press_Schechter_z3_261223_PK.eps}}
%{\includegraphics{UV_Press_Schechter_z3_271223_PK.eps}}
%{\includegraphics{UV_Press_Schechter_z3_281223_PK.eps}}
%{\includegraphics{UV_Press_Schechter_z3_291223_PK.eps}}
{\includegraphics{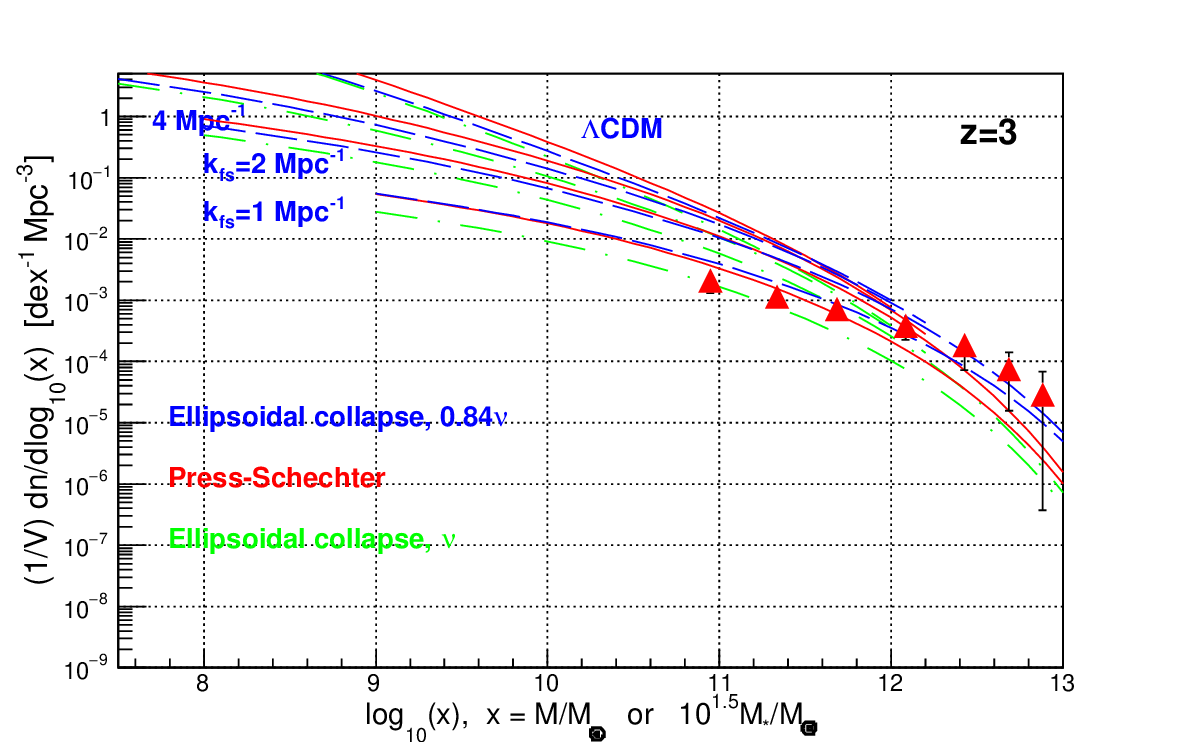}}
\scalebox{0.33}
%{\includegraphics{UV_Press_Schechter_z3_091223_.eps}}
%{\includegraphics{UV_Press_Schechter_z3_111223_.eps}}
%{\includegraphics{UV_Press_Schechter_z3_211223_.eps}}
%{\includegraphics{UV_Press_Schechter_z3_261223_.eps}}
%{\includegraphics{UV_Press_Schechter_z3_271223_.eps}}
%{\includegraphics{UV_Press_Schechter_z3_281223_.eps}}
{\includegraphics{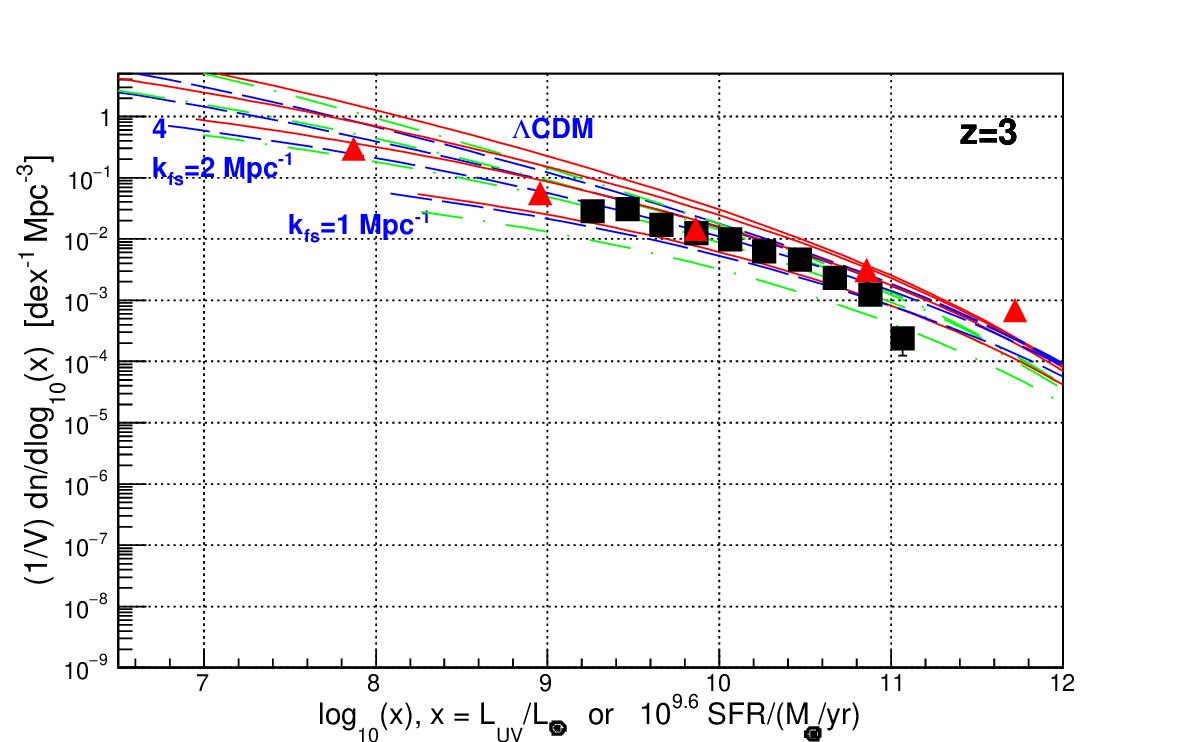}}
\scalebox{0.33}
%{\includegraphics{UV_Press_Schechter_z4_091223_PK.eps}}
%{\includegraphics{UV_Press_Schechter_z4_101223_PK.eps}}
%{\includegraphics{UV_Press_Schechter_z4_111223_PK.eps}}
%{\includegraphics{UV_Press_Schechter_z4_211223_PK.eps}}
%{\includegraphics{UV_Press_Schechter_z4_281223_PK.eps}}
%{\includegraphics{UV_Press_Schechter_z4_311223_PK.eps}}
{\includegraphics{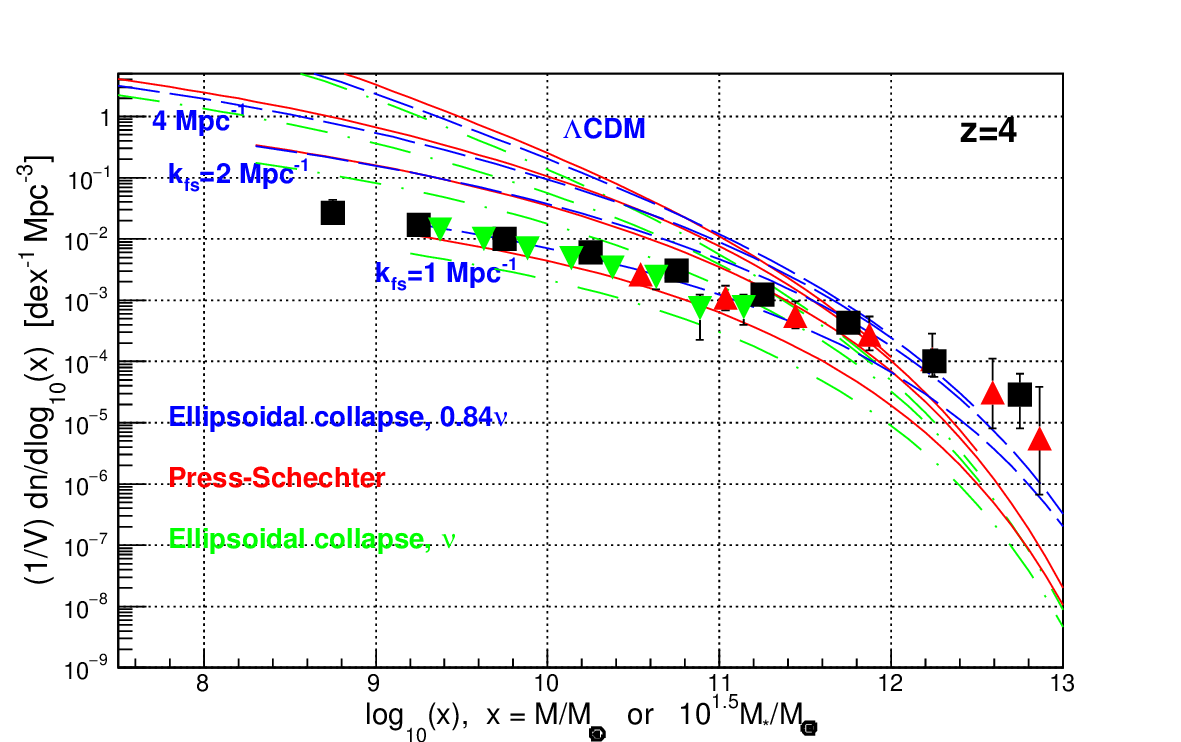}}
\scalebox{0.33}
%{\includegraphics{UV_Press_Schechter_z4_091223_.eps}}
%{\includegraphics{UV_Press_Schechter_z4_111223_.eps}}
%{\includegraphics{UV_Press_Schechter_z4_211223_.eps}}
%{\includegraphics{UV_Press_Schechter_z4_281223_.eps}}
%{\includegraphics{UV_Press_Schechter_z4_311223_.eps}}
{\includegraphics{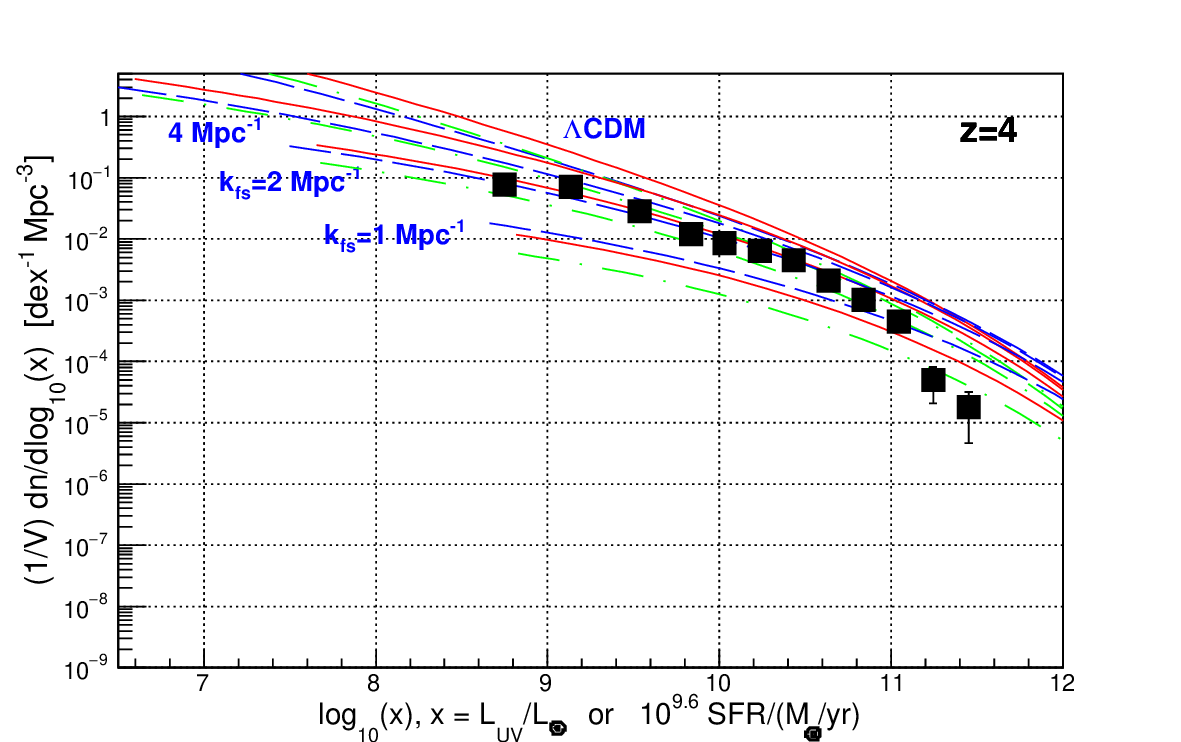}}
%\vspace*{0.7cm}
\caption{Comparison of predicted and observed
distributions of $M/M_\odot = 10^{1.5} M_*/M_\odot$ (left panels) 
and $L_{UV}/L_\odot = 10^{9.6} \textrm{SFR}/(M_\odot/\textrm{yr})$
(right panels) for redshift $z = 2$ (top row), 3 and 4
(bottom row). 
Data are from the Hubble Space Telescope (black squares) ($M_*$ from \cite{Song} and
$L_{UV}$ from \cite{Bouwens}),
from the continuity equation \cite{Lapi} (red triangles), and
from the James Webb Space Telescope (green triangles) \cite{Navarro}.
%https://iopscience.iop.org/article/10.3847/0004-637X/825/1/5/pdf \cite{Song}
%https://arxiv.org/pdf/2102.07775.pdf \cite{Bouwens}
%https://arxiv.org/pdf/1708.07643.pdf \cite{Lapi}
}
\label{PS_z2-4}
%/home/bruce1/JWST
%UV_Press_Schechter.C_bck091223
\end{center}
\end{figure}

\begin{figure}
\begin{center}
%\vspace*{-4.5cm}
\scalebox{0.33}
%{\includegraphics{UV_Press_Schechter_z5_091223_PK.eps}}
%{\includegraphics{UV_Press_Schechter_z5_111223_PK.eps}}
%{\includegraphics{UV_Press_Schechter_z5_211223_PK.eps}}
%{\includegraphics{UV_Press_Schechter_z5_281223_PK.eps}}
{\includegraphics{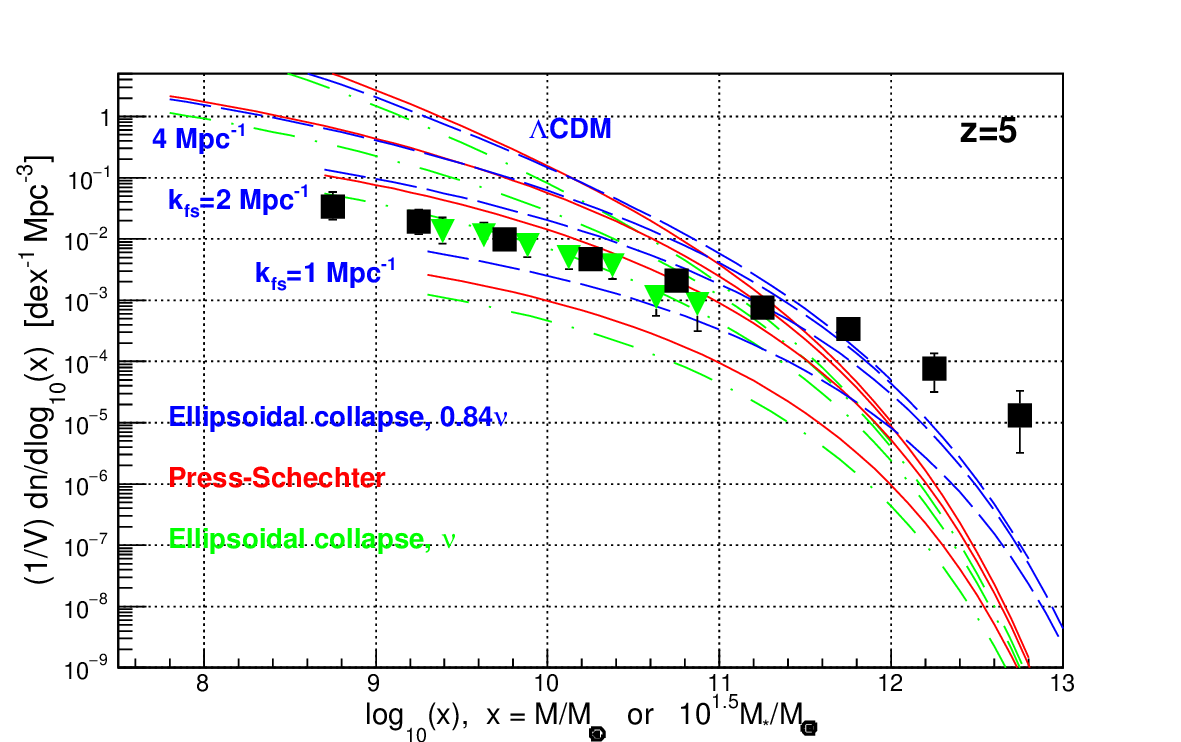}}
\scalebox{0.33}
%{\includegraphics{UV_Press_Schechter_z5_091223_.eps}}
%{\includegraphics{UV_Press_Schechter_z5_111223_.eps}}
%{\includegraphics{UV_Press_Schechter_z5_211223_.eps}}
%{\includegraphics{UV_Press_Schechter_z5_281223_.eps}}
{\includegraphics{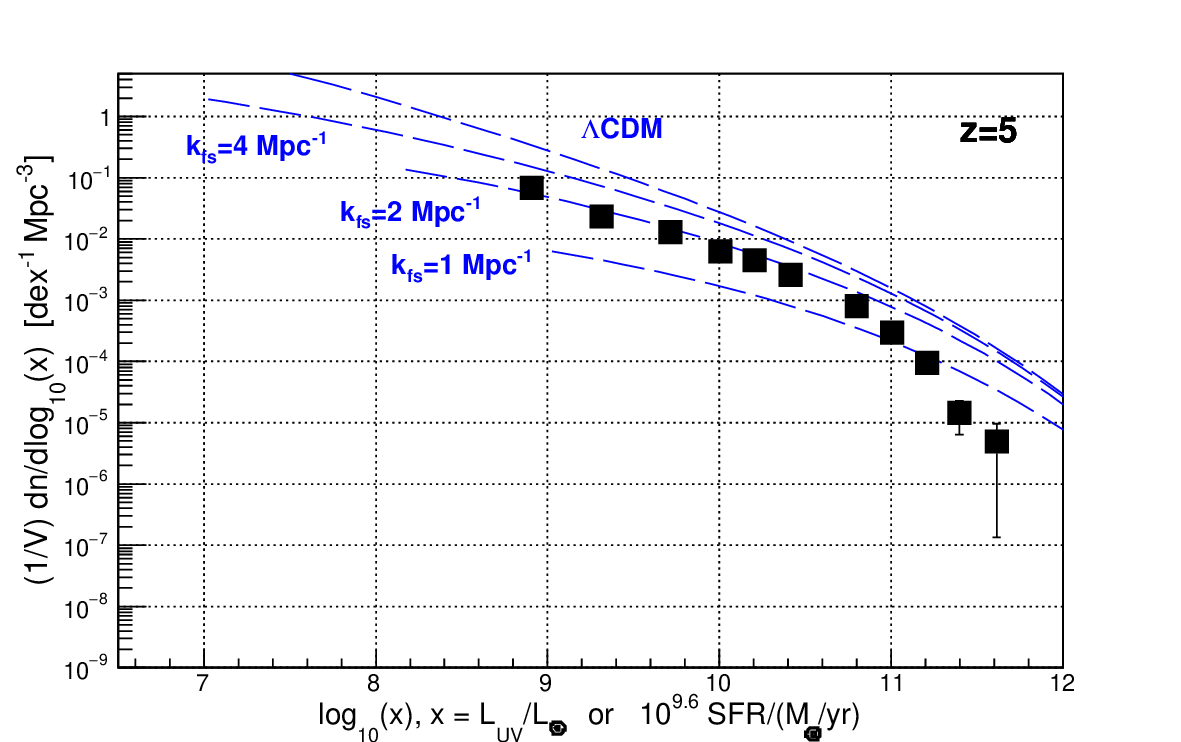}}
\scalebox{0.33}
%{\includegraphics{UV_Press_Schechter_z7_091223_PK.eps}}
%{\includegraphics{UV_Press_Schechter_z7_111223_PK.eps}}
%{\includegraphics{UV_Press_Schechter_z7_281223_PK.eps}}
%{\includegraphics{UV_Press_Schechter_z7_311223_PK.eps}}
{\includegraphics{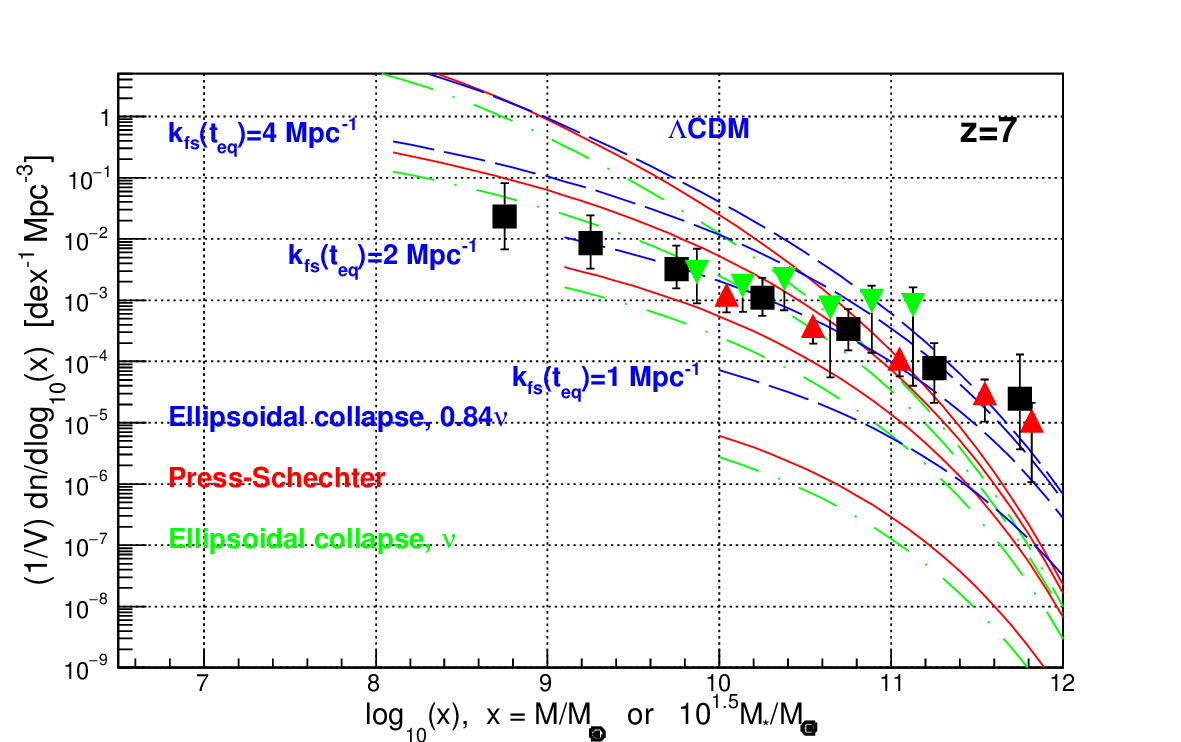}}
\scalebox{0.33}
%{\includegraphics{UV_Press_Schechter_z7_091223_.eps}}
%{\includegraphics{UV_Press_Schechter_z7_111223_.eps}}
%{\includegraphics{UV_Press_Schechter_z7_281223_.eps}}
%{\includegraphics{UV_Press_Schechter_z7_311223_.eps}}
{\includegraphics{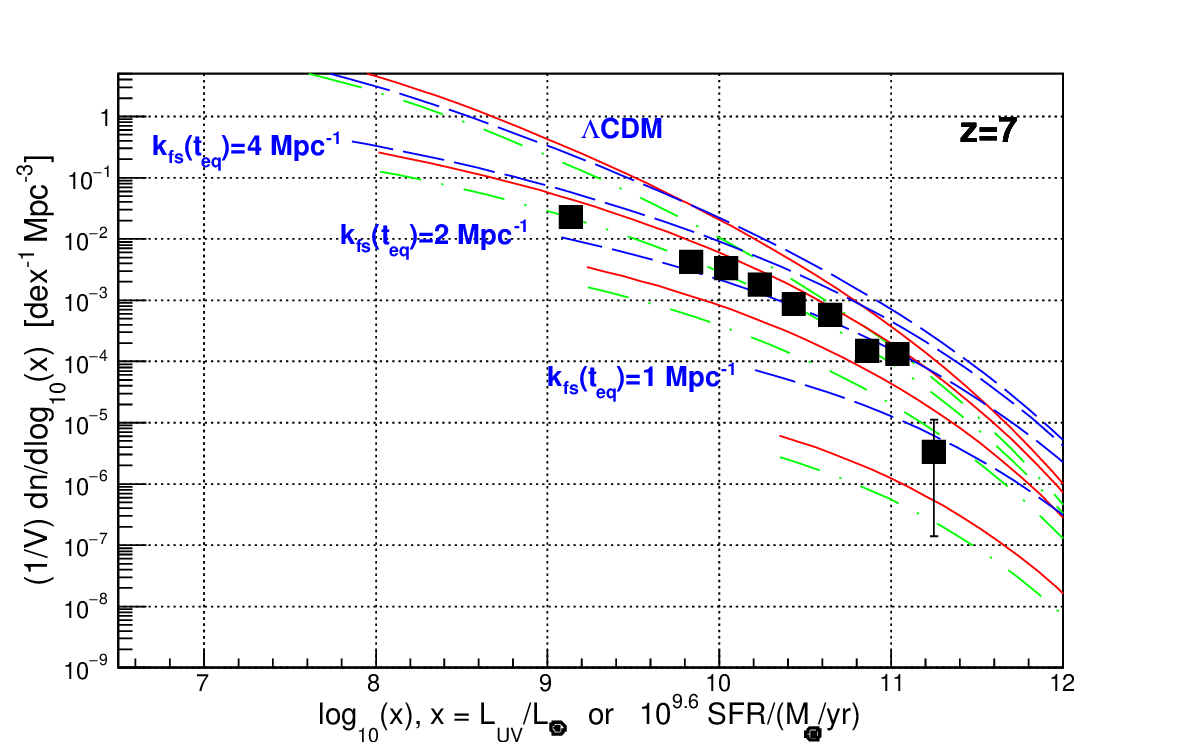}}
\scalebox{0.33}
%{\includegraphics{UV_Press_Schechter_z8_081223_PK.eps}}
%{\includegraphics{UV_Press_Schechter_z8_091223_PK.eps}}
%{\includegraphics{UV_Press_Schechter_z8_101223_PK.eps}}
%{\includegraphics{UV_Press_Schechter_z8_111223_PK.eps}}
%{\includegraphics{UV_Press_Schechter_z8_261223_PK.eps}}
%{\includegraphics{UV_Press_Schechter_z8_281223_PK.eps}}
%{\includegraphics{UV_Press_Schechter_z8_311223_PK.eps}}
{\includegraphics{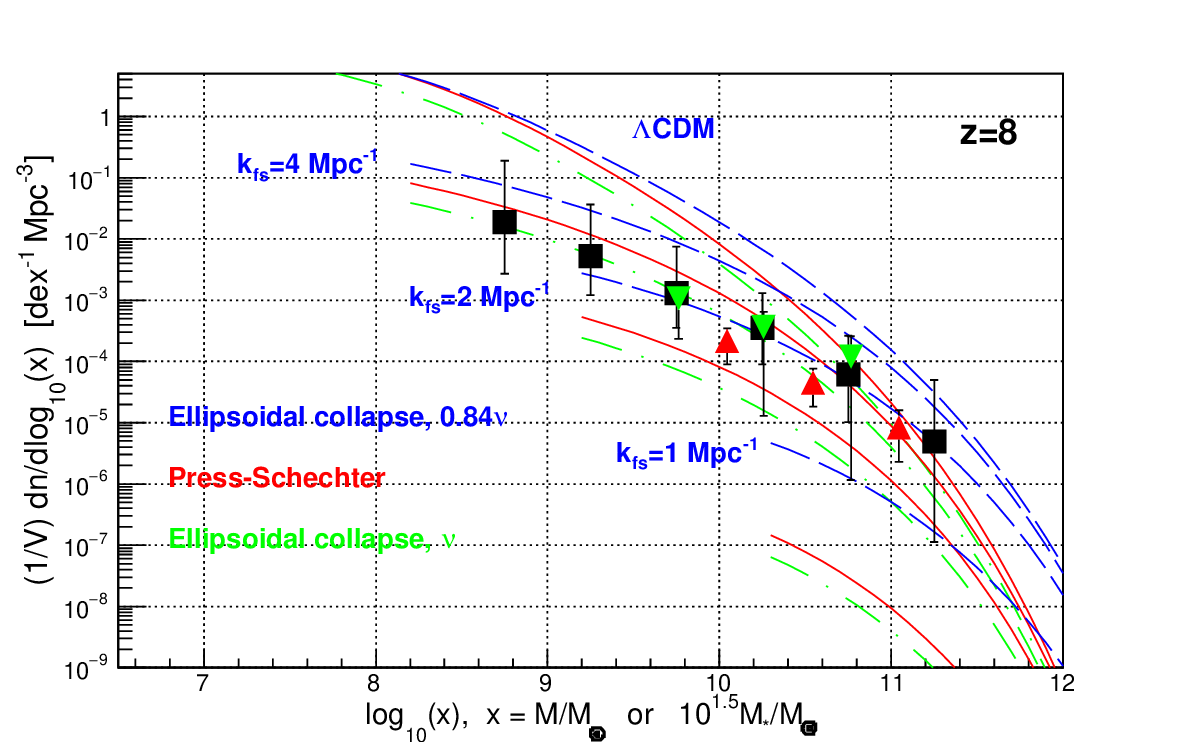}}
\scalebox{0.33}
%{\includegraphics{UV_Press_Schechter_z8_081223_.eps}}
%{\includegraphics{UV_Press_Schechter_z8_091223_.eps}}
%{\includegraphics{UV_Press_Schechter_z8_111223_.eps}}
%{\includegraphics{UV_Press_Schechter_z8_261223_.eps}}
%{\includegraphics{UV_Press_Schechter_z8_281223_.eps}}
%{\includegraphics{UV_Press_Schechter_z8_311223_.eps}}
{\includegraphics{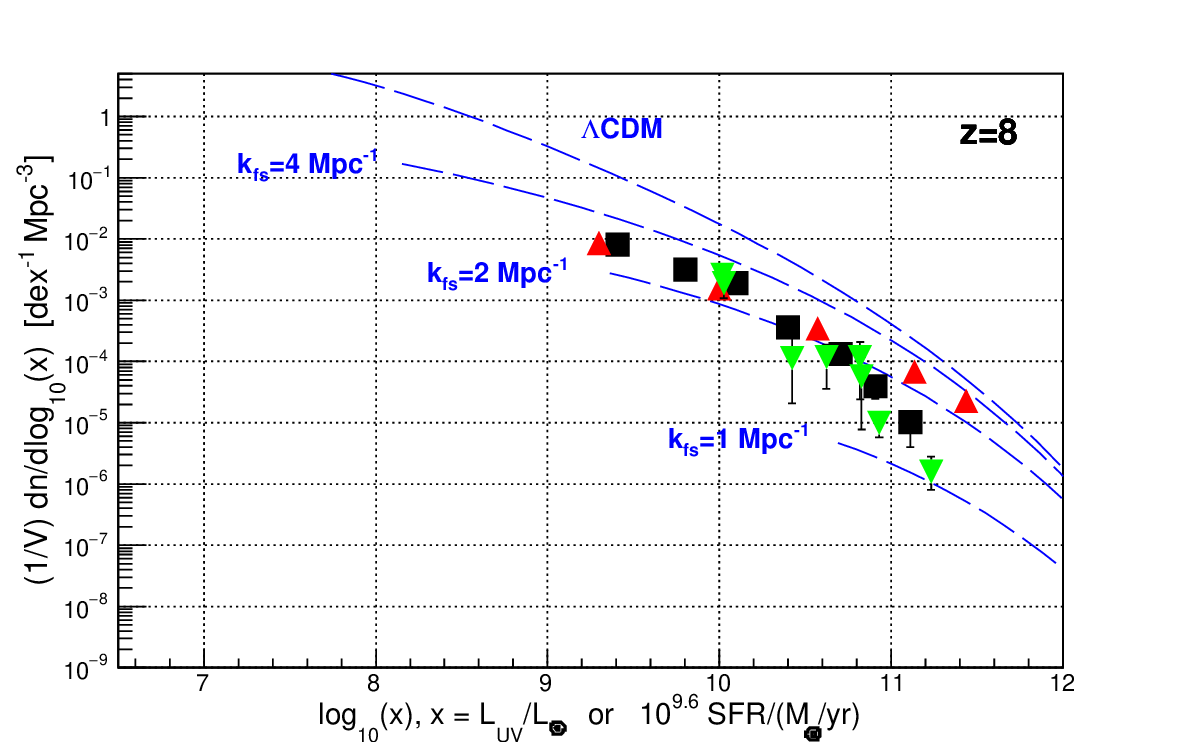}}
%\vspace*{0.7cm}
\caption{Comparison of predicted and observed
distributions of $M/M_\odot = 10^{1.5} M_*/M_\odot$ (left panels) 
and $L_{UV}/L_\odot = 10^{9.6} \textrm{SFR}/(M_\odot/\textrm{yr})$
(right panels) for redshift $z = 5$ (top row), 7 and 8
(bottom row).
Data are from the Hubble Space Telescope (black squares) ($M_*$ from \cite{Song} and
$L_{UV}$ from \cite{Bouwens}),
from the continuity equation \cite{Lapi} (red triangles), and
from the James Webb Space Telescope (green triangles) ($M_*$ from \cite{Navarro} and
$L_{UV}$ from \cite{Adams23}, \cite{Bouwens23a}, \cite{Donnan23}).
%https://iopscience.iop.org/article/10.3847/0004-637X/825/1/5/pdf \cite{Song}
%https://arxiv.org/pdf/2102.07775.pdf \cite{Bouwens}
%https://arxiv.org/pdf/1708.07643.pdf \cite{Lapi}
}
\label{PS_z5-8}
%/home/bruce1/JWST
%UV_Press_Schechter.C_bck091223
\end{center}
\end{figure}

\begin{figure}
\begin{center}
%\vspace*{-4.5cm}
\scalebox{0.33}
%{\includegraphics{UV_Press_Schechter_z9_091223_PK.eps}}
{\includegraphics{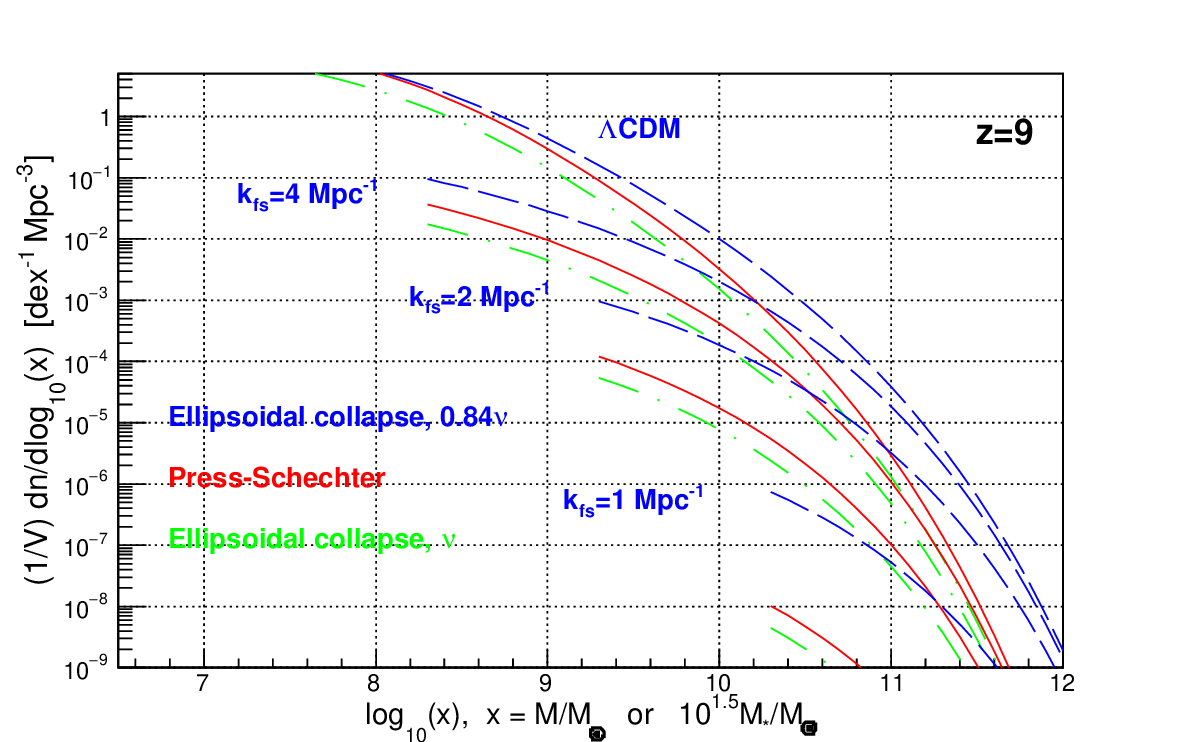}}
\scalebox{0.33}
%{\includegraphics{UV_Press_Schechter_z9_091223_.eps}}
{\includegraphics{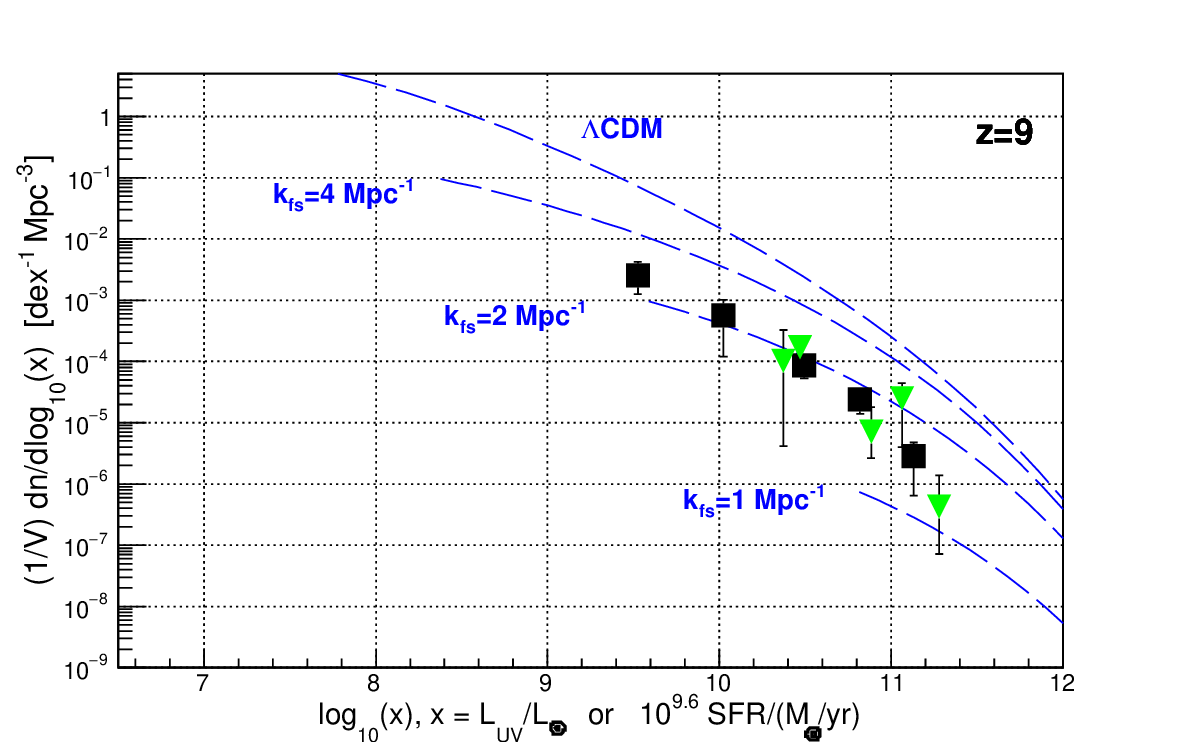}}
\scalebox{0.33}
%{\includegraphics{UV_Press_Schechter_z10_081223_PK.eps}}
{\includegraphics{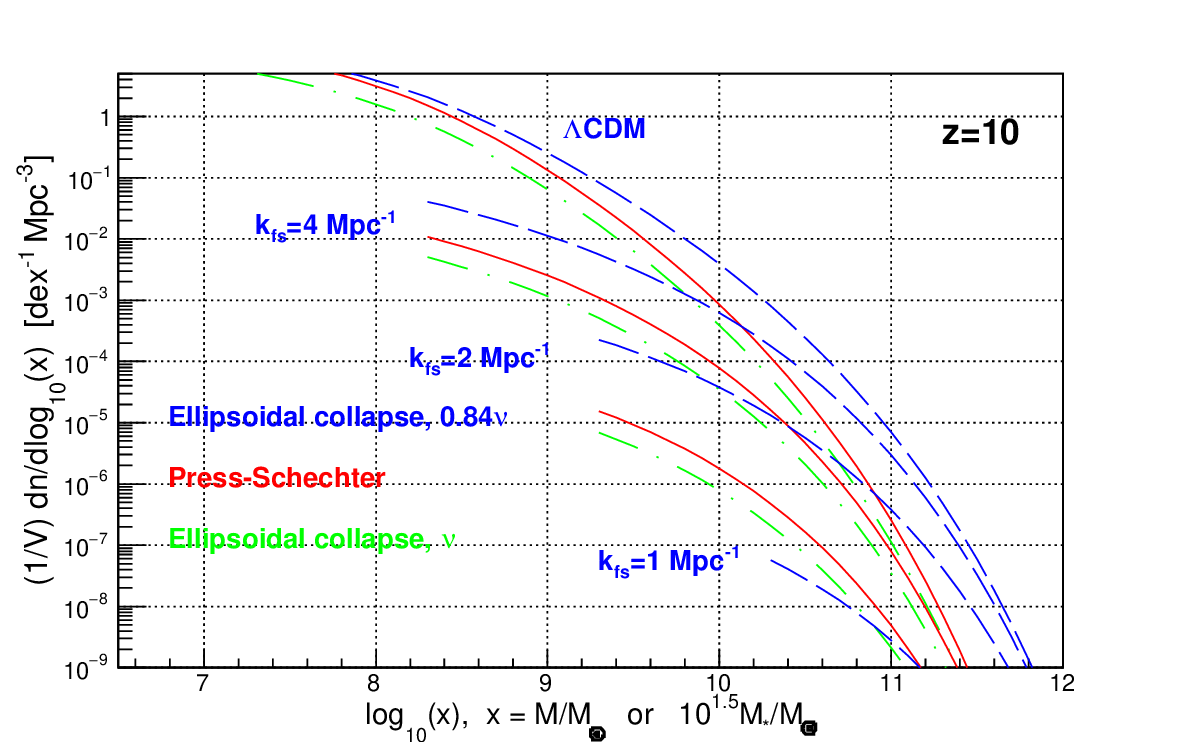}}
\scalebox{0.33}
%{\includegraphics{UV_Press_Schechter_z10_081223_.eps}}
{\includegraphics{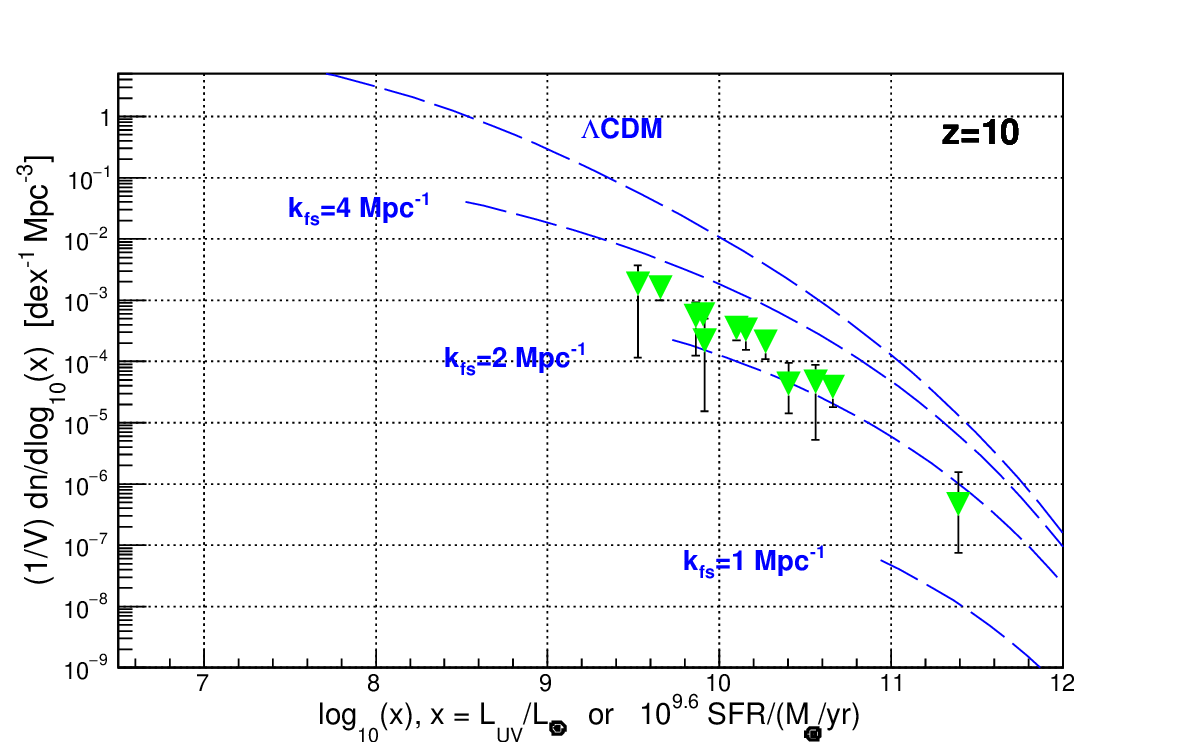}}
\scalebox{0.33}
%{\includegraphics{UV_Press_Schechter_z11_081223_PK.eps}}
%{\includegraphics{UV_Press_Schechter_z11_281223_PK.eps}}
{\includegraphics{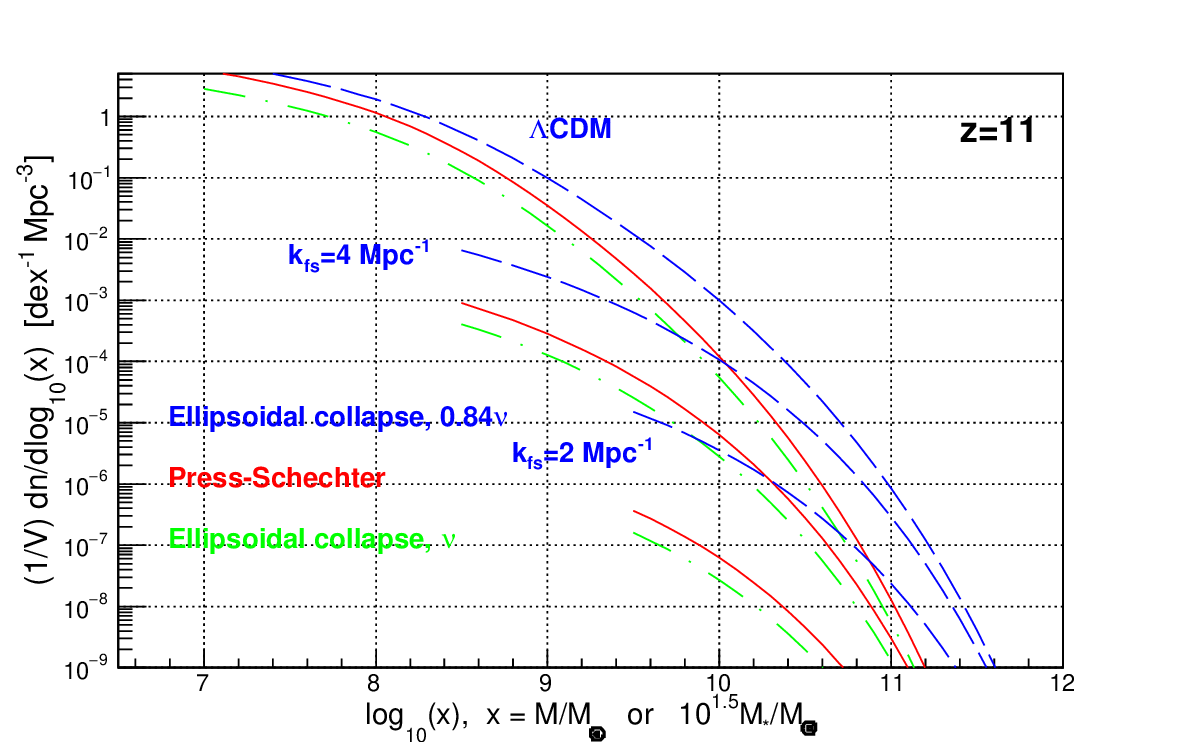}}
\scalebox{0.33}
%{\includegraphics{UV_Press_Schechter_z11_081223_.eps}}
%{\includegraphics{UV_Press_Schechter_z11_281223_.eps}}
{\includegraphics{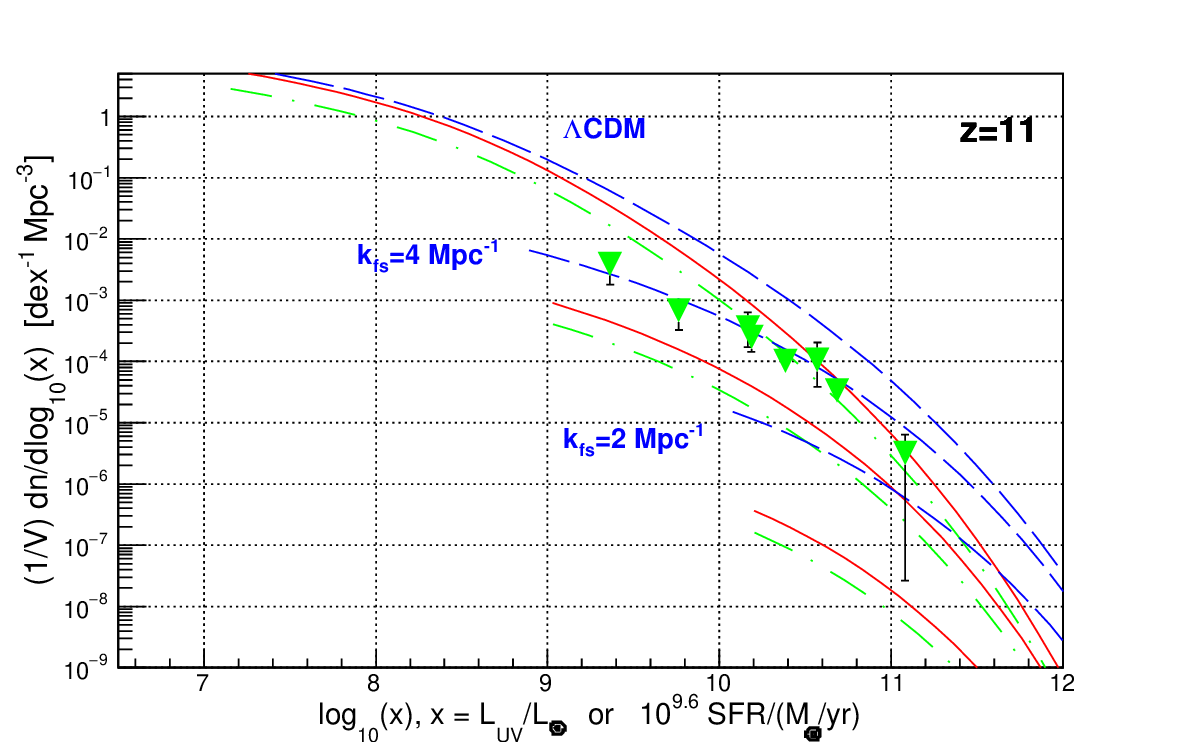}}
%\vspace*{0.7cm}
\caption{Comparison of predicted and observed
distributions of $M/M_\odot = 10^{1.5} M_*/M_\odot$ (left panels with no data) 
and $L_{UV}/L_\odot = 10^{9.6} \textrm{SFR}/(M_\odot/\textrm{yr})$
(right panels) for redshift $z = 9$ (top row), 10 and 11
(bottom row).
Data are from the Hubble Space Telescope (black squares)
($L_{UV}$ from \cite{Bouwens}), and
from the James Webb Space Telescope (green triangles) 
($L_{UV}$ from \cite{Adams23}, \cite{Bouwens23a}, \cite{Donnan23}, \cite{Harikane23b},
\cite{Finkelstein23}, \cite{Mcleod23}, \cite{Perez23}).
%Fig. 3 of arxiv.2307.12487 \cite{Bouwens}
%Table 2 of https://iopscience.iop.org/article/10.3847/0004-637X/825/1/5/pdf \cite{Song}
%\cite{Lapi}
}
\label{PS_z9-11}
%/home/bruce1/JWST
%UV_Press_Schechter.C_bck091223
\end{center}
\end{figure}

\begin{figure}
\begin{center}
%\vspace*{-4.5cm}
\scalebox{0.33}
%{\includegraphics{UV_Press_Schechter_z12_081223_PK.eps}}
{\includegraphics{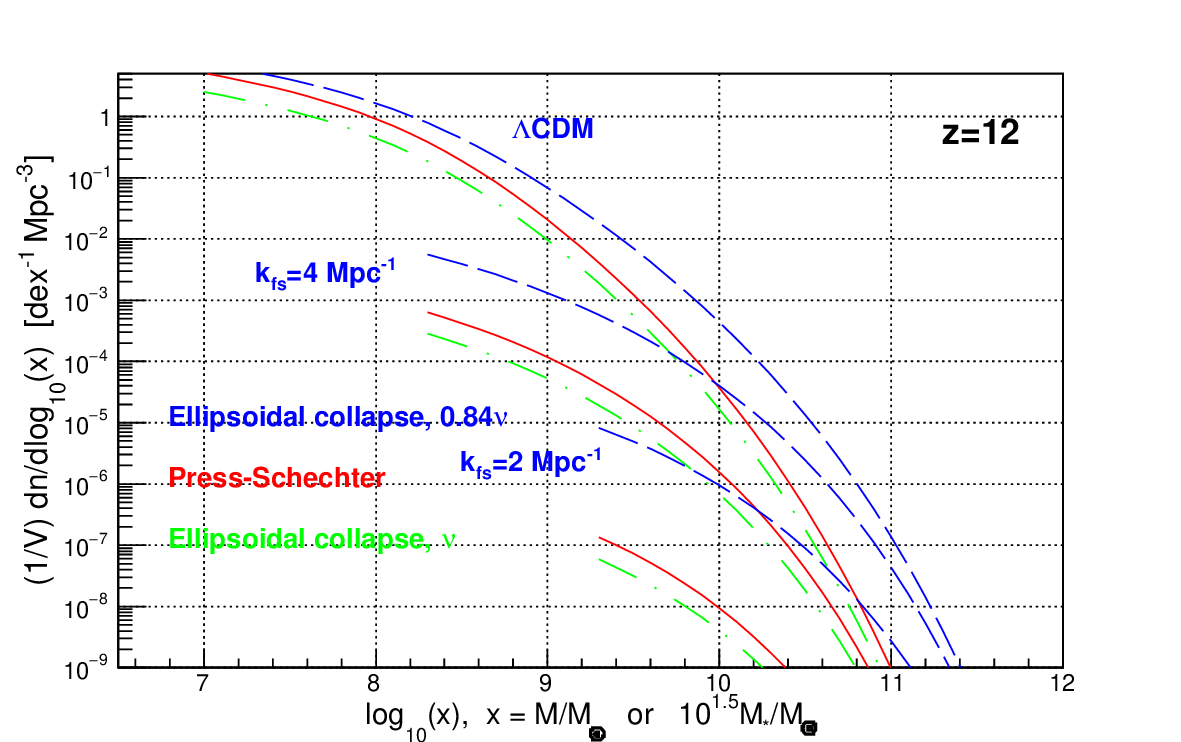}}
\scalebox{0.33}
%{\includegraphics{UV_Press_Schechter_z12_081223_.eps}}
{\includegraphics{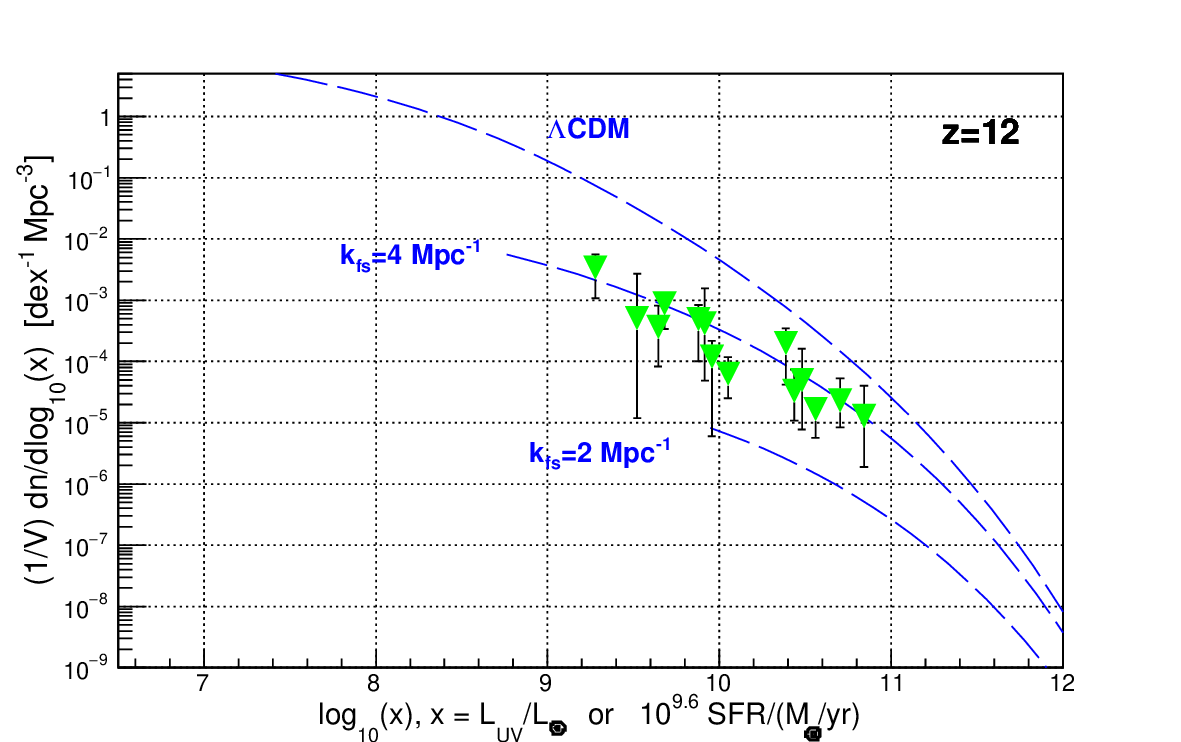}}
\scalebox{0.33}
%{\includegraphics{UV_Press_Schechter_z13_081223_PK.eps}}
%{\includegraphics{UV_Press_Schechter_z13_091223_2_PK.eps}}
{\includegraphics{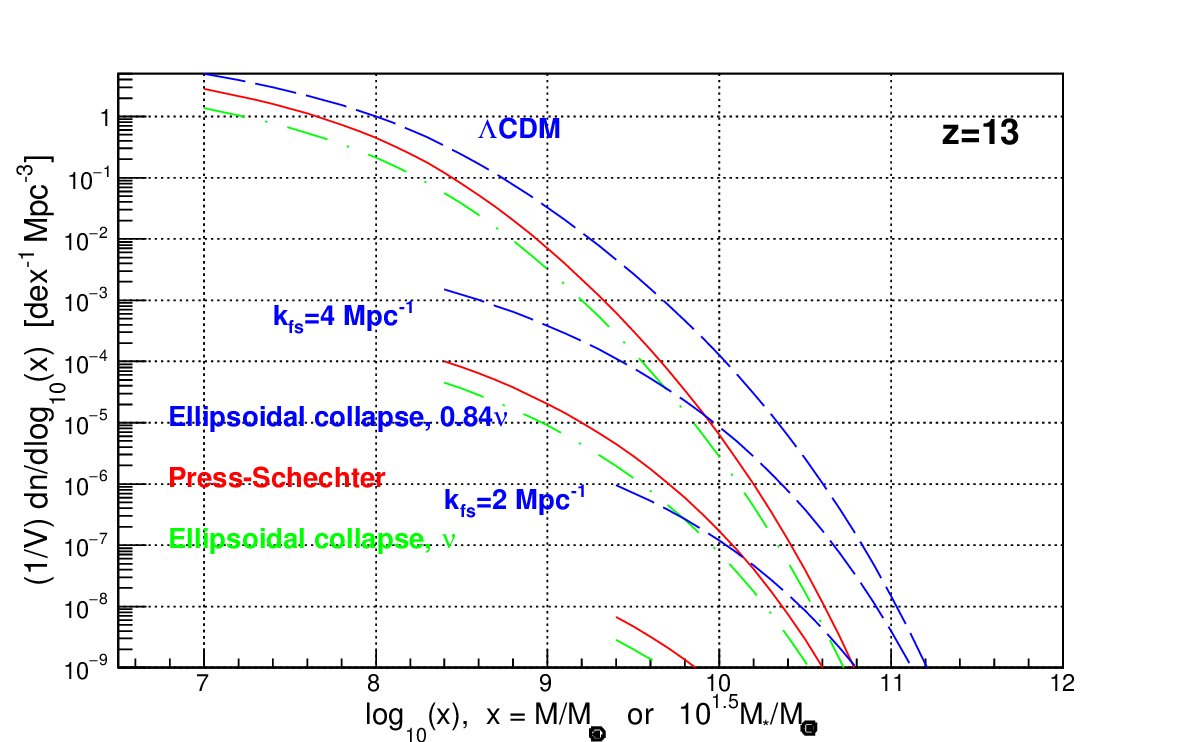}}
\scalebox{0.33}
%{\includegraphics{UV_Press_Schechter_z13_081223_.eps}}
%{\includegraphics{UV_Press_Schechter_z13_091223_2.eps}}
{\includegraphics{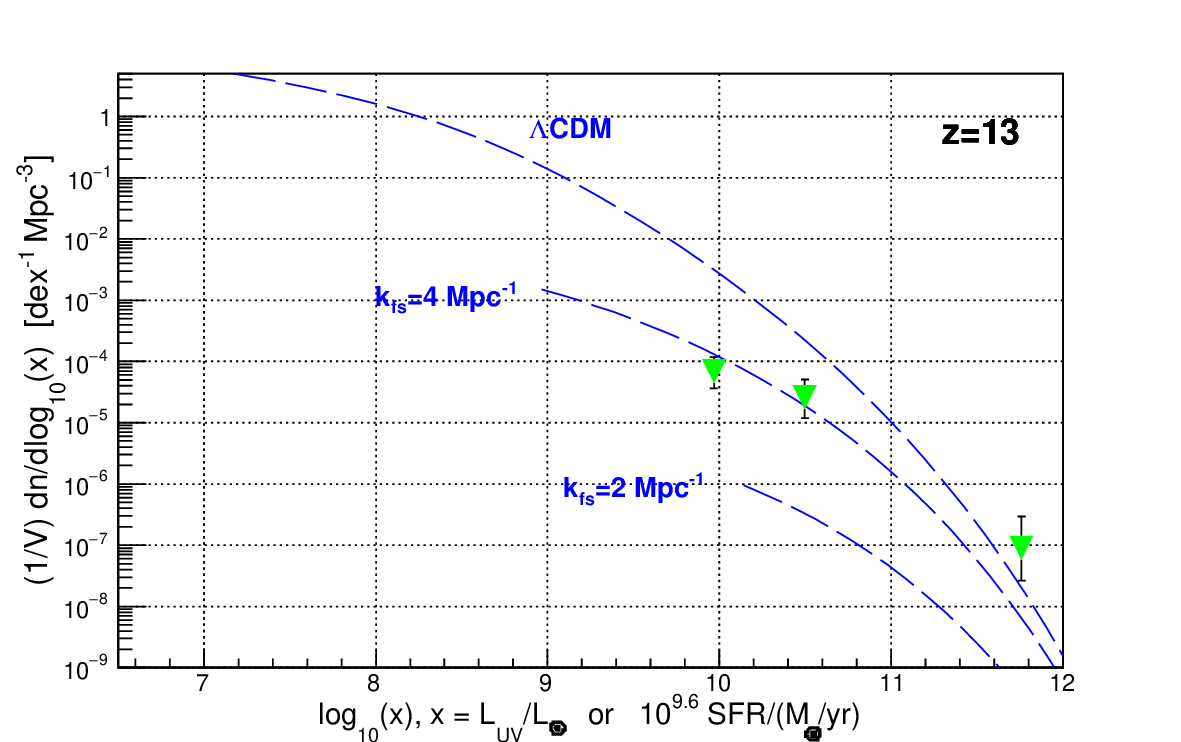}}
\scalebox{0.33}
{\includegraphics{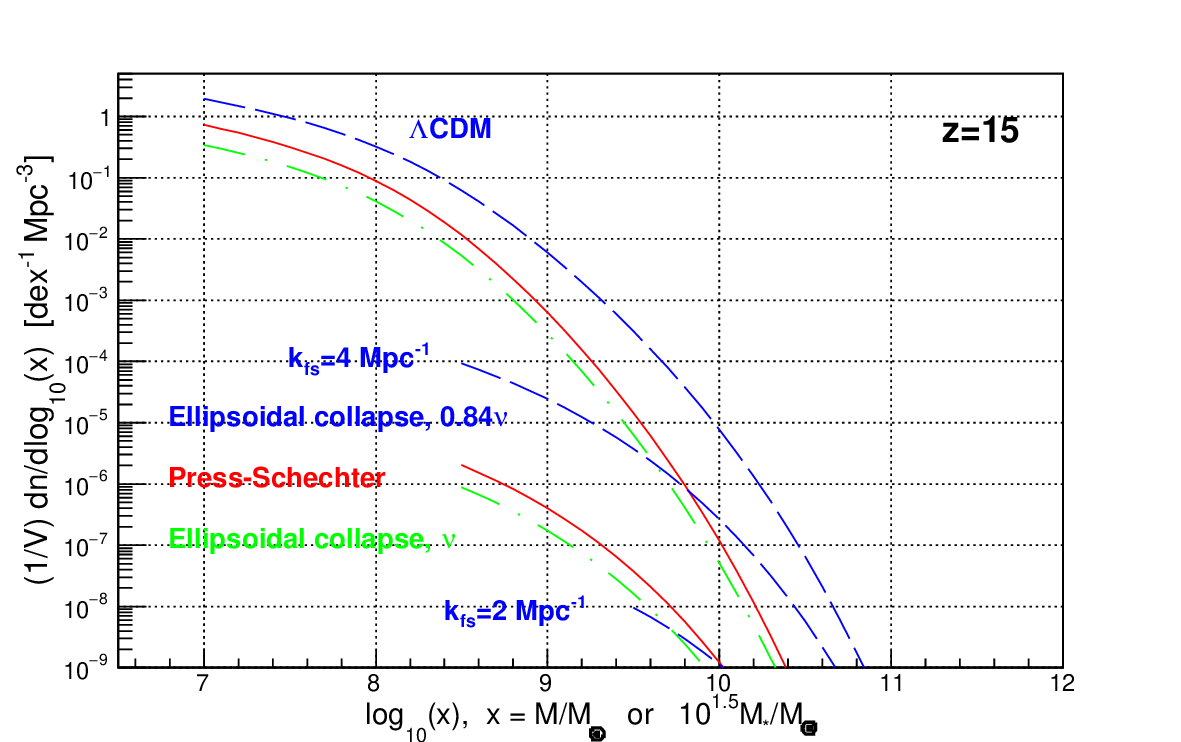}}
\scalebox{0.33}
{\includegraphics{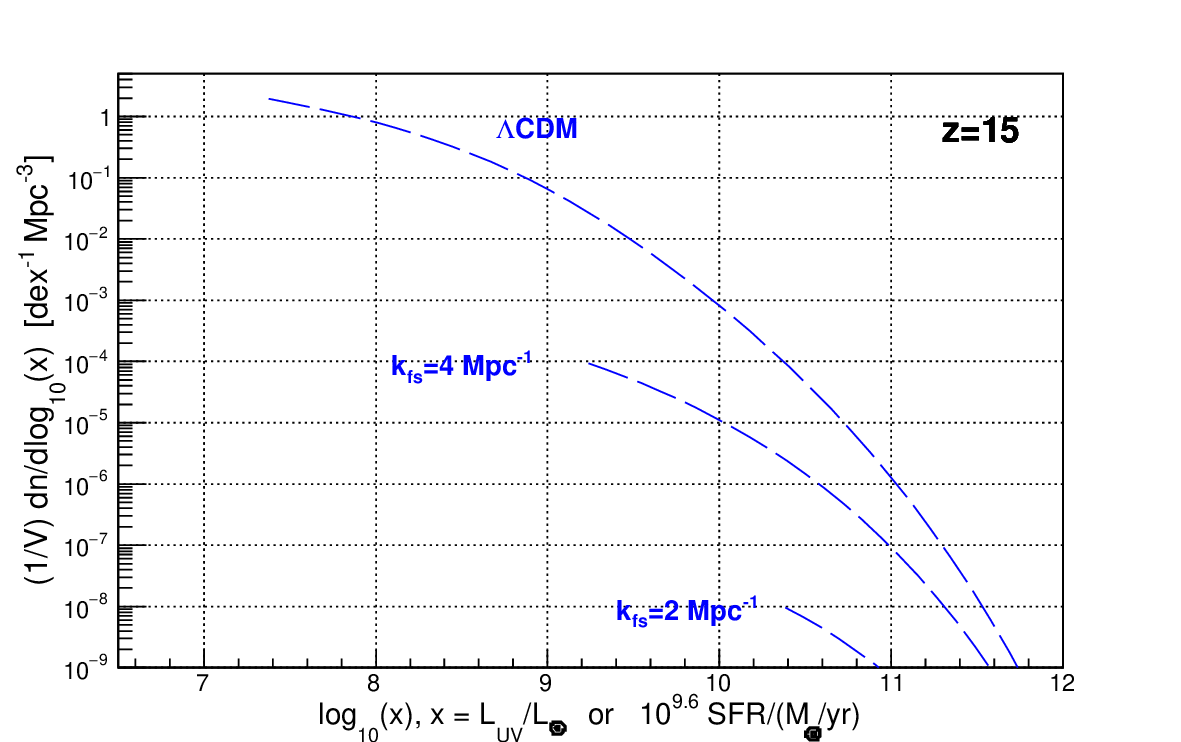}}
%\vspace*{0.7cm}
\caption{Comparison of predicted and observed
distributions of $M/M_\odot = 10^{1.5} M_*/M_\odot$ (left panels with no data) 
and $L_{UV}/L_\odot = 10^{9.6} \textrm{SFR}/(M_\odot/\textrm{yr})$
(right panels) for redshift $z = 12$ (top row), 13 and 15
(bottom row).
Data are from the James Webb Space Telescope (green triangles) 
($L_{UV}$ from 
\cite{Adams23}, \cite{Bouwens23a}, \cite{Donnan23}, \cite{Harikane23b},
\cite{Finkelstein23}, \cite{Perez23}, \cite{Morishita23}, \cite{Naidu22}).
%Fig. 3 of arxiv.2307.12487 
}
\label{PS_z12-15}
%/home/bruce1/JWST
%UV_Press_Schechter.C_bck081223
\end{center}
\end{figure}

Comparisons of observed distributions of $M_*/M_\odot$ and $L_{UV}/L_\odot$ 
with first-order predictions, for redshift $z$ in the range 2 to 13, 
are presented in Figures \ref{PS_z6} to \ref{PS_z12-15}.
Sometimes we omit the ``Press-Schechter" and ``Ellipsoidal Collapse, $\nu$" predictions for clarity.
For future reference we also present predictions for $z = 15$ in Figure \ref{PS_z12-15}.
The predictions are presented for $M \gtrsim M_{vd}$, where $M_{vd}$ is the
velocity dispersion limit of validity of the predictions \cite{LUV}:
for $M \lesssim M_{vd}$ the density fluctuation does not collapse gravitationally.
We find agreement of the measurements with the first-order predictions
in most of the parameter space, so these predictions are a good
starting point to develop a more detailed theory.

There are however three discrepant regions:
\begin{enumerate}
\item
Discrepancies of $L_{UV}$ distributions at $L_{UV}/L_\odot \gtrsim 10^{10.7}$ and
$3 \lesssim z \lesssim 9$.
\item
Discrepancies of $M_*$ distributions at $z \lesssim 5$.
\item
Discrepancies with preliminary JWST observations of $L_{UV}$ distributions at $z \gtrsim 11$.
\end{enumerate}
The first and second discrepancies are common to cold and warm dark matter.

We will discuss these discrepancies in Section \ref{discussion}.
However, before doing so, we need to understand the data sources,
and the warm dark matter extension of the theory.

\section{Data}
\label{data}

For $z = 2$ to 10 we obtain the distributions of magnitude $M_{1600,AB}$ from
Table 4 of \cite{Bouwens} (for $z = 10$ the results are from \cite{Oesch}). 
These measurements have black square markers
in Figures \ref{PS_z6} to \ref{PS_z9-11}.
The data in these references are obtained from Hubble Space Telescope (HST) observations
in approximately 11 filter bands (to obtain the photometric $z$ with
stellar population synthesis (SPS) models),
i.e. $UV_{275}, B_{435}, V_{606}, z_{850}, V_{606}, UV_{336},
I_{814}, J_{125}$, $Y_{105}, H_{160}, JH_{140}$,
and from Spitzer Space Telescope (SST) observations.

For $z = 4$ to 8 we obtain the distributions of galaxy stellar mass $M_*/M_\odot$ from
Table 2 or Figure 9 of \cite{Song} (identified by
black square markers in distributions of $M_*/M_\odot$). 
The analysis in \cite{Song} obtains relations between $M_*/M_\odot$ and
magnitude $M_{UV}$, for each photometric $z$, fitting HST images
taken with 10 filters, using an SPS model.
Then the distributions of $M_*/M_\odot$ are obtained with the
distributions of magnitude $M_{UV}$ in \cite{Finkelstein}.
As a cross-check we mention that the thick gray lines in Figure 9 of 
\cite{Song}, corresponding to $\Lambda$CDM
predictions, are in agreement with our $\Lambda$CDM predictions. 

We obtain distributions of galaxy stellar mass $M_*/M_\odot$ for $z = 2$ to 8 from
Figure 4 of \cite{Lapi}, and distributions of star formation rates (SFR)
from Figure 1 of \cite{Lapi} (red triangles).
These distributions are obtained using the continuity equation for the stellar masses of galaxies,
with inputs and comparisons to a large number of sources \cite{Lapi}.

For $z = 4$ to 8 we obtain the distributions of galaxy stellar mass $M_*/M_\odot$ from
Figure 5 of \cite{Navarro} (green triangles). These measurements are based on approximately
3300 galaxy images taken with the James Webb Space Telescope
Near Infrared Camera (NIRCam).
This data is complemented by the HST Cosmic Assembly
Near-Infrared Deep Extragalactic Legacy Survey (CANDELS).
The label ``$z = 6$" in, e.g. Figure \ref{PS_z6}, means $z = 6$ for the predictions,
and $5.5 < z < 6.5$ for the measurements. We have neglected the corresponding bias.

Measurements of $L_{UV}$ for $z \ge 8$ from JWST data are obtained from 
\cite{Adams23}, \cite{Bouwens23a}, \cite{Donnan23},
\cite{Harikane23b}, \cite{Finkelstein23}, \cite{Mcleod23}, 
\cite{Perez23}, \cite{Morishita23}, \cite{Naidu22}
(red triangles). 
See \cite{Wang} for an analysis in the $\Lambda$CDM scenario.
The measurements for $z \gtrsim 11$ need to be considered as ``preliminary", as
stated by their authors, for several reasons:
\begin{itemize}
\item
The measurements at lower $z$ are counting experiments: the galaxy
candidates are assigned to bins of magnitude and redshift $(M_{UV}, z)$
and are then counted. 
At the high-$z$ frontier, the bins $(M_{UV}, z)$ are increased in size
so the mean number of counts does not drop below $\approx 1$. The analysis
ceases to be a counting experiment, and becomes an extrapolation of a
Press-Schechter-like fit to lower $z$ data, keeping all parameters fixed
except the normalization.
The tension arises at the transition between these two methods.
\item
To illustrate the low numbers of events, let us mention that the
measurements in \cite{Adams23} are based on 33, 22, 16 and 3 galaxy candidates,
after correcting for completeness, in the bins
$7.5 < z < 8.5$, $8.5 < z < 9.5$, $9.5 < z < 11.5$ and
$11.5 < z < 13.5$, respectively.
Therefore, where the
apparent tension between observations and ``first-order"
predictions of $L_{UV}$ arise, i.e. $z \gtrsim 11$,
there are only about 3 galaxy candidates with photometric redshift.
On the other hand, a single galaxy correctly assigned to a bin, i.e. with spectroscopically confirmed
redshift $z$, and a luminosity distribution consistent with stellar
expectations (limiting the possible contribution from active galactic nuclei (AGN)), 
could rule out a theory if the theory predicts one galaxy with 
a probability less than, say, 0.6\%.
\item
To illustrate the difficulties with the photometric classification of galaxies
let us mention that Table 4 of \cite{Bouwens23a} presents an assessment of the
``purity" and ``completeness" of the selected samples by several authors.
A good feeling of the uncertainties is quoted from \cite{Bouwens23a}: ``Using all
of these samples we then derive UV LF and luminosity density
results at $z \ge  8$, finding substantial differences. For example,
including the full set of “solid” and “possible” $z \ge 12$
candidates from the literature, we find UV LF and luminosity densities which
are $\approx 7\times$ and $\approx 20\times$ higher than relying on the “robust” candidates
alone. These results indicate the evolution of the UV LF and
luminosity densities at $z \ge 8$ is still extremely uncertain,
emphasizing the need for spectroscopy and deeper NIRCam$+$optical
imaging to obtain reliable results."
Note that the measurements of $L_{UV}$ in Figures \ref{PS_z9-11} and \ref{PS_z12-15} 
are based on photometric redshift measurements.
\item
If AGN contribute to the observed UV luminosity,
then the measurements can drop by 0.4 dex on average, or up to
4 dex for individual galaxies \cite{DSilva}.
\item
Corrections for dust attenuation are very uncertain unless data is
available in a wide range of wavelengths \cite{Lapi}.
\item
The measurements of $L_{UV}$ presented in Figures \ref{PS_z5-8}
to \ref{PS_z12-15} for $z \ge 8$ are not independent as they use
overlapping data sets. At $z \ge 11$ there are more measurements than galaxy candidates.
\item
Cosmic variance becomes important at $z \gtrsim 14$ \cite{Yung}.
\end{itemize}

The experimental determinations of the galaxy stellar masses $M_*$ 
depend on stellar population synthesis (SPS) models and 
spectral energy distributions (SED) models that use images with 
several filters, and therefore do not 
include dead star remnants (ejected baryons into inter-stellar space, white dwarfs,
neutron stars, and black holes). 
So measured and predicted $M_*$ do not include dead star remnants.

\section{Predictions}
\label{predictions}

This section describes the extension of the Press-Schechter
formalism that we use to include warm dark matter (see \cite{LUV}
and \cite{comments} for more details).
The Press-Schechter prediction \cite{Press-Schechter} is
\begin{equation}
\frac{1}{V} \frac{d n}{d \ln{M}} = \frac{\bar{\rho}_m}{M} \frac{d \ln (\sigma^{-1})}{d \ln M} f_\textrm{PS}(\nu),
\label{PS}
\end{equation}
where
\begin{equation}
f_\textrm{PS}(\nu) = \sqrt{ \frac{2}{\pi} } \nu \exp{\left(- \frac{\nu^2}{ 2} \right)},
\label{fPS}
\end{equation}
and 
\begin{equation}
\nu \equiv \frac{1.686}{\sigma(M, z, k_{fs})}.
\label{nu}
\end{equation}
The factor 1.686 is obtained analytically for spherical collapse with cold dark matter,
and becomes valid for warm dark matter when $M \gtrsim M_{vd}$, where $M_{vd}$ is the
velocity dispersion cut-off mass. 
Predictions are presented only for $M > M_{vd}$, see Table 1 of \cite{LUV}.
In the spirit of the present study, we
define our ``first-order" prediction, for comparison purposes, with the factor 1.686 unchanged, 
i.e. independent of $M$ and $k_{fs}$.
The Sheth-Mo-Tormen ellipsoidal collapse extensions \cite{Sheth_Tormen} \cite{Sheth_Mo_Tormen} are
obtained by replacing $f_\textrm{PS}(\nu)$ by $f_\textrm{EC}(\nu)$:
\begin{equation}
f_\textrm{EC}(\nu) = 0.322 \left[ 1 + \tilde{\nu}^{-0.6} \right] f_\textrm{PS}(\tilde{\nu}),
\label{fEC}
\end{equation}
with $\tilde{\nu} = \nu$.
Good fits to simulations are obtained with
$\tilde{\nu} = 0.84 \nu$ \cite{Sheth_Mo_Tormen}. The factor 0.84 depends on
the algorithm used to identify the collapsed halos.

These predictions depend on the variance of the linear relative
density perturbation $\delta(\textbf{x}) \equiv (\rho(\textbf{x}) - \bar{\rho})/\bar{\rho}$:
\begin{equation}
\sigma^2(M, z, k_\textrm{fs}) = \frac{f^2}{(2 \pi)^3 (1 + z)^2} \int_0^\infty 4 \pi k^2 dk P(k) \tau^2(k) W^2(k).
\end{equation}
$\sigma^2(M, z, k_\textrm{fs})$ depends on the linear total (dark matter plus baryon) 
mass scale $M$ at redshift $z$, and on the comoving cut-off wavenumber $k_{fs}$ due
to dark matter free-streaming. $P(k)$ is the comoving power spectrum of linear density
perturbations in the cold dark matter $\Lambda$CDM cosmology \cite{Weinberg}. $W(k)$ is a window function
that defines the mass scale $M$. $\tau^2(k)$ is the power spectrum 
cut-off factor due to dark matter free-streaming.
The factor $f$ is due to the cosmological constant: $f = 0.79, 0.99, 1.00$ for
$z = 0, 2, \gg 2$ \cite{Weinberg}.
Three window functions are considered: the top-hat in r-space, the sharp-k (or top-hat in k-space),
and the Gaussian window function:
\begin{equation}
W(k) = \exp{\left( -\frac{k^2}{2 k_0^2} \right)}, \qquad
M = \frac{4}{3} \pi \left( \frac{1.555}{k_0} \right)^3 \bar{\rho}_m.
\end{equation}
$\bar{\rho}_m$ is the total (dark matter plus baryon) mean density.
We use the following form for the free-streaming cut-off factor:
\begin{eqnarray}
\tau^2(k) & = & \exp{\left( -\frac{k^2}{k^2_\textrm{fs}(t_\textrm{eq})} \right)} \qquad
\textrm{ if } k < k_\textrm{fs}(t_\textrm{eq}), \nonumber \\
& = & \exp{\left( -\frac{k^n}{k^n_\textrm{fs}(t_\textrm{eq})} \right)}\qquad
\textrm{   if } k \ge k_\textrm{fs}(t_\textrm{eq}).
\label{tail}
\end{eqnarray}
At the time $t_{eq}$ when the matter density begins to dominate, $\tau^2(k)$ has the 
approximate form (\ref{tail}) with $n = 2$ \cite{Boyanovsky}.
Thereafter, $\tau^2(k)$ develops a non-linear regenerated ``tail" with
$n$ measured to be in the range 0.5 to 1.1 \cite{LUV}.
In the present study we take $n = 1$, and use the Gaussian window function
(see studies in \cite{LUV}).
We have verified that the predictions change negligibly for $0.5 < n < 1.1$,
and also if the sharp-k window function is used with $n = 1$ \cite{LUV}.
The amplitude of $P(k)$ is adjusted for each $k_{fs}$ so that the relative density
root-mean-square fluctuation $\sigma$, calculated with 
the top-hat window function with radius $r = 8/h = 8/0.674$ Mpc,
is $\sigma_8 = 0.811$ \cite{PDG2023}.

\section{Discussion}
\label{discussion}

Let us now discuss the discrepancies found in Figures \ref{PS_z6} to \ref{PS_z12-15}. 

1. Discrepancies of $L_{UV}$ distributions at $L_{UV}/L_\odot \gtrsim 10^{10.7}$ and
$3 \lesssim z \lesssim 9$. \\
This discrepancy is observed in HST and JWST data.
This discrepancy is studied in \cite{Lapi}, and is (apparently)
due to the dust	correction of the measured SFR.	When the dust correction
is based only on the UV slope, the results are inconsistent with
other data sets, and the dust-corrected SFR falls short	of
multi-wavelength determinations at high SFR (dominated by dusty 
star forming progenitors of present-day quiescent galaxies).
A reliable dust correction needs, in addition to the UV images,
also optical, radio, H$\alpha$, mid-IR 24 $\mu$m, and far-IR
images.	The results of the continuity equation \cite{Lapi},
based on multi-wavelength dust corrections, are indeed consistent with the
predictions as shown in Figures \ref{PS_z6} to \ref{PS_z5-8} (see the 
red triangles in $L_{UV}$ distributions). 

2. Discrepancies of $M_*$ distributions at $z \lesssim 5$. 
This discrepancy can be understood qualitatively, at least in part, as follows
(this is my tentative understanding).
In the warm dark matter	scenario, the first galaxies to form
have $M	\approx	M_{vd} \approx 2 \times	10^8 M_\odot$ at $z = 4$,
increasing to $\approx 2 \times 10^9 M_\odot$ at $z = 10$ (see Table 1 of \cite{LUV} for $k_{fs} = 2$ Mpc$^{-1}$).
Thereafter, the formation of galaxies proceeds hierarchically as
larger and larger perturbation masses $M$ become non-linear and collapse due to gravity.
Therefore, low mass halos become part of higher and higher
mass halos as time goes on, and so become under-counted.
In other words, the distribution of low $M_*$ galaxies becomes
ill-defined at late times.

3. Discrepancies of preliminary JWST observations of $L_{UV}$ distributions at $z \gtrsim 11$. 
For $6 \lesssim z \lesssim 10$ the JWST data is in agreement with predictions
for $k_{fs} \approx 2$ Mpc$^{-1}$
(see green triangles in Figures \ref{PS_z6} to \ref{PS_z9-11}).
Preliminary JWST observations at $z = 11, 12$ and 13 are in mild tension
with the predictions. However these observations need spectroscopic
confirmation of the redshifts, and higher statistics, 
before any conclusions can be presented,
see discussion in Section \ref{data} and in \cite{Bouwens23a}.

\section{Conclusions}
\label{conclusions}

We have	presented comparisons of measured distributions	of
galaxy luminosities per unit bandwidth $L_{UV}$ and
galaxy stellar masses $M_*$ with ``first-order" predictions,
as a function of $k_{fs}$,
for $z$	in the range 2 to 13.
The only outstanding tension for JWST observations corresponds to
$L_{UV}$ with $z \gtrsim 11$. However, these measurements are still
preliminary until spectroscopic	confirmation of	$z$ is obtained,
and of limited significance (as stated by the authors, e.g. \cite{Bouwens23a})

We conclude that the
``first-order" predictions with the measured $k_{fs} = 2.0^{+0.8}_{-0.5}$ Mpc$^{-1}$ \cite{LUV},
and two	parameters $a =	1.5$ and $b = 9.6$, obtained \textit{prior} to 
JWST data \cite{fermion_or_boson} \cite{LUV}, and assumed to be constants
independent of $(M, z, k_{fs})$, are in agreement with
the current data within their theoretical and observational 
uncertainties. This result is indeed surprising,
considering the large range of $z$ and $M$, with constant star formation efficiency $f_*$
and constant $b$, and such a basic and simple ``first order" prediction model
with no new degrees of freedom!
The ``first-order" predictions are therefore a useful starting point to include
more detailed astrophysical models to account for more
precise	future observations, and to describe other observables, and their evolution.

Can dark matter	be cold? Consider, as an example, Figure \ref{PS_z6}.
The distributions of $M_*/M_\odot$ and of $L_{UV}/L_\odot$
are nicely consistent with $k_{fs} \approx 2$ Mpc$^{-1}$ and a
constant star formation efficiency $f_* \approx 0.20$.
If, however, dark matter is cold, i.e. if $k_{fs}$ is very large, then
both $M_*$ and $L_{UV}$ data tells us that
$f_*$ at $M \approx 10^9 M_\odot$ is approximately $1/10$th of $f_*$ at $M \gtrsim 10^{11} M_\odot$.
Is this	possible? Supernova and active galactic nuclei (AGN) winds are invoked
to explain this low $f_*$ at low $M_*$. However,
three additional and independent indications that 
dark matter may be warm, with $k_{fs} \approx 2$ Mpc$^{-1}$,
are i) the observed distributions of $M = 10^{1.5} M_*$ have a cut-off at approximately $M_{vd}$
(less massive galaxies would have to be
``stripped-down" galaxies \cite{comments}), ii) the measurements of spiral galaxy rotation
curves \cite{formation} \cite{dwarf}, and iii) the re-ionization optical depth
\cite{LUV} \cite{Lin}. 
Item i) is related to the ``missing satellite" problem, and
ii) is related to the ``core-cusp" problem.

If the preliminary tension of $L_{UV}$ with $z \gtrsim 11$
is confirmed by	future observations and	analysis, and becomes significant, 
then we need to let
the parameters $a$ and $b$ become functions of $z$,
with $b - a$ growing from 8.1 at $z \lesssim 10$ to        
approximately 9.1 at $z \gtrsim 11$ (for the case $k_{fs} \approx 2$ Mpc$^{-1}$). 
In other words, we would have to allow the
star formation efficiency $f_*$	to increase above 0.20 at $z \gtrsim 11$,
and/or allow first stars to be more massive and
luminous than at lower $z$. Let us work and see.

\end{document}